\documentclass[12pt]{article}
\usepackage{jheppub}
\newcommand{\intd}{\mathrm{d}}
\newcommand{\ex}{\mathrm{e}}

\newcommand{\hp}[1]{\hphantom{#1}}
\usepackage{amsmath,epsfig}
\usepackage{amssymb,amsfonts}
\usepackage{latexsym}
\usepackage{slashed}
\usepackage{empheq}
\numberwithin{equation}{section}
\usepackage{dsfont}
\usepackage{cancel}
\usepackage{overpic}
\usepackage{subcaption}
\usepackage{bibtopic}
\usepackage{subeqnarray}
\usepackage{xcolor}

\usepackage{graphicx}
\usepackage{amsmath}
\usepackage{longtable}
\usepackage{color}

\allowdisplaybreaks

\newcommand{\exclude}[1]{}
\def\nn{\nonumber}

\def\L{\mathcal{L}}

\def\d{\mathrm{d}}

\def\a#1{\alpha_{#1}}

\def\beq{\begin{equation}}
	\def\eeq{\end{equation}}
\def\be{\begin{equation}}
	\def\ee{\end{equation}}
\def\bea{\begin{eqnarray}}
	\def\eea{\end{eqnarray}}
\def\bal{\begin{align}}
	\def\eal{\end{align}}

\def\Im{\textrm{Im}}

\def\2b2[#1,#2][#3,#4]{\left( \begin{array}{cc} #1 & #2 \\ #3 & #4 \end{array}
	\right)}
\def\3b3[#1,#2,#3][#4,#5,#6][#7,#8,#9]{\left( \begin{array}{ccc} #1 & #2 #3 \\
		#4 & #5 & #6\\#7&#8&#9\end{array} \right)}

\newcommand\fverb{\setbox\pippobox=\hbox\bgroup\verb}
\newcommand\fverbdo{\egroup\medskip\noindent%
	\fbox{\unhbox\pippobox}\ }
\newcommand\fverbit{\egroup\item[\fbox{\unhbox\pippobox}]}

\newcommand{\bear}{\begin{eqnarray}}
	
	\newcommand{\eear}{\end{eqnarray}}

\newcommand{\bsea}{\begin{subeqnarray}}
	\newcommand{\esea}{\end{subeqnarray}}
\newbox\pippobox

\def\d{\delta}

\def\6{\partial}
\def\a{\alpha}
\def\nn{\nonumber}

\def\pa{\partial}

\def\e{\epsilon}
\def\m{\mu}
\def\n{\nu}

\def\s{\sigma}

\def\sp{\;\;\;,\;\;\;}

\def\sq
\def\a{\alpha}
\def\b{\beta}

\def\hri#1#2{\href{http://arxiv.org/abs/#1}{[ArXiv:#1]#2}}
\def\hre#1#2{\href{http://arxiv.org/abs/#1/#2}{[ArXiv:#1/#2]}}
\def\hrj#1#2{\href{https://doi.org/#1}{#2}}

\def\e{\epsilon}

\def\d{\delta}

\def\L{\Lambda}

\def\D{\Delta}

\def\EE{{\cal E}}

\def\OO{{\cal O}}

\title{Revisiting color superconductors in bottom-up holography}
\author{
	E. Pr\'eau}

\affiliation{Institute for Theoretical Physics, Utrecht University, 3584 CE Utrecht, The Netherlands}

\preprint{}
\abstract{We revisit the problem of constructing a color superconducting phase in bottom-up holography.  We introduce a model that describes the five-dimensional dynamics of a scalar field dual to a chiral symmetry breaking condensate, which is coupled to the RR 4-form. The coupling is given by a topological WZ term, which is elegantly responsible both for scalar condensation at high density, and color Higgsing in the condensed phase. This construction realizes both properties at finite density and Higgsing fraction.
}

\begin{document}
\maketitle

\section{Introduction}
\label{Intro}

After more than five decades of theoretical and experimental progress, there is little doubt that the strong interaction can be described by the theory of quantum chromodynamics (QCD). However, many properties of QCD remain poorly understood, due to the strongly-coupled nature of the theory \cite{Brambilla:2014jmp}. Among these, one of the most prominent challenges is to understand the phase diagram of QCD at finite baryonic density \cite{Fukushima:2010bq}. 

In this direction, important progress was made from the realization that many properties of QCD matter can be computed at densities much larger than the confining scale, for which QCD becomes weakly coupled. In this regime, perturbative QCD predicts the dominant phase to correspond to a so-called color superconductor (CSC) \cite{Rajagopal:2000wf,Alford:2001dt,Alford:2007xm}, where the formation of quark pairs induces color Higgsing. Several types of CSC phases were found to exist, among which the color-flavor locked (CFL) phase is expected to be of particular phenomenological relevance \cite{Rajagopal:2000wf,Alford:2001dt,Alford:2007xm}. Such a CFL phase occurs for $N_c=3$ colors and $N_f=3$ light flavors of quarks, and is characterized by the spontaneous breaking of chiral symmetry and baryon number.

As the density is lowered, it is expected that the CSC regime should connect to the low density nuclear matter phase, which may happen through a phase transition or a continuous crossover \cite{Alford:2001dt}. However, answering this question is a hard problem, since QCD becomes strongly-coupled in the regime of intermediate densities (a few times the QCD scale). In this part of the phase diagram, no first-principle approach can be used to compute the properties of matter, including the numerical lattice QCD method which is made ineffective by the sign problem \cite{Troyer:2004ge}.

In the last thirty years, an alternative analytic tool has emerged to study strongly-coupled theories, provided  by the holographic correspondence \cite{Malda,GKP,Witten98,review}. The latter makes it possible to solve (large N) non-perturbative problems, by performing instead classical calculations in a dual gravitational theory, which is defined on a higher-dimensional space-time (the bulk).

Several holographic models of QCD have been constructed \cite{Sakai:2004cn,Sakai:2005yt,Kruczenski:2003be,Erlich:2005qh,DaRold:2005mxj,Karch:2006pv,VQCDvac}, that share many properties with QCD, such as confinement, similar hadronic spectra, flavor anomalies and chiral symmetry breaking. The most sophisticated of these models \cite{VQCDvac}, are defined over the full parameter space at finite temperature \cite{VQCDT} and density \cite{VQCDmu,Jokela:2018ers,Demircik:2021zll}, and give fairly faithful descriptions of QCD in regimes where data is available. 

However, color superconductivity is an aspect of QCD that is hard to realize in holography. The main obstacle faced by holographic models, is that the naive CSC ``order parameters" are not gauge invariant, so that they do not admit dual bulk fields. 

Nevertheless, there do exist holographic realizations of CSC phases, which are based on top-down duals\footnote{See also \cite{BitaghsirFadafan:2018iqr,BitaghsirFadafan:2020otb,Ghoroku:2019trx} for a bottom-up phenomenological approach based on the assumption that the color group can be treated as a global symmetry group.} of supersymmetric cousins of QCD\footnote{See \cite{Harnik:2003ke} for a field theoretic analysis of color superconductivity in superQCD.} \cite{Apreda:2005yz,Chen:2009kx,Faedo:2018fjw}. These theories possess a single trace gauge-invariant order parameter for CSC phases, which is given by the di-squark condensate. The constructions of \cite{Apreda:2005yz,Chen:2009kx} realize the dynamical generation of a CSC phase at finite temperature and small density, whereas that of \cite{Faedo:2018fjw} corresponds to a CSC phase at finite isospin density. An important limitation of these top-down setups, is that they are only tractable in the regime where the number of Higgsed colors is much smaller than the total $N_c$. 

In this work, we consider instead a bottom-up model for a similar QCD-like theory, in which di-squark condensation induces color Higgsing at high density. At the cost of introducing indeterminacy in the model, the bottom-up approach makes it possible to circumvent certain technical difficulties of top-down models. In particular, the regime of large Higgsing, as well as that of high density, are naturally accessible for appropriate model parameters. 

From the bottom-up point of view, it is not straightforward how color Higgsing can be realized in the five-dimensional bulk theory. A general method that provides useful insight for bottom-up constructions, is to consider the dimensional reduction of top-down settings. In string-theoretic models, color Higgsing is realized by the color branes merging into the flavor branes\footnote{In the case of $\mathcal{N}=4$ SYM, Higgsing may also be realized without any fields in the fundamental representation (which defines the so-called Coulomb branch). The bulk dual of such configurations contains explicit D3 branes \cite{Malda}, that are placed in the geometry generated by the unbroken color group. Reference \cite{Henriksson:2022mgq} describes a bottom-up model that is designed to reproduce this kind of $\mathcal{N}=4$ Higgsed phase.}. The fused color branes then appear as instantons for the worldvolume gauge fields, that wrap the internal part of the geometry \cite{Apreda:2005yz,Chen:2009kx,Faedo:2018fjw}. When projected to five dimensions, the instantons reduce to a worldvolume scalar field\footnote{See appendix \ref{AppA1} for more details.}, which has the right properties to be the dual of the di-squark operator \cite{Faedo:2018fjw}.

Following these lessons from top-down setups, our model is constructed to describe the five-dimensional dynamics of the di-squark scalar field, which is also responsible for color Higgsing via its coupling to the Ramond-Ramond (RR) color flux. The coupling is given by a Wess-Zumino (WZ) term whose form is fixed by RR gauge invariance, and can also be seen to arise from the projection of top-down models (see appendix \ref{AppA1}). Elegantly, this WZ coupling is crucial both for triggering di-squark condensation at high density, and the associated color Higgsing. The resulting CSC solution fully accounts for the back-reaction of all fields, at finite Higgsing fraction. 

The organization of this paper is as follows. In the rest of this introduction, we provide a more detailed summary of our results, followed by an outlook. In section \ref{S1}, we describe how the holographic model is constructed, while section \ref{S2} contains the analysis of stability to di-squark condensation, together with the description of the phase diagram. Section \ref{S3} finally discusses the properties of the CSC phase, including the dependence on model parameters. Additional details are provided in the appendices. 

\subsection{Summary of results}
\label{I3}

Our holographic model describes the dynamics of a scalar $\chi$ dual to the di-squark operator\footnote{It may also be seen as the dual of the usual chiral condensate $\bar{\psi}\psi$. Only the scalar UV mass distinguishes the two cases.}, which is coupled in the flavor sector to a $U(1)_B$ baryon number gauge field $A_\mu$, and in the color sector to the RR 4-form $C_4$ and gravity. 

The bulk action is the sum of three terms 
\begin{equation}
\nn S=S_c+S_f+S_{WZ} \, ,
\end{equation}
that are constructed by taking inspiration from string theoretic settings, as discussed in section \ref{S1}. The color part of the action $S_c$ contains the kinetic terms for gravity (Einstein-Hilbert) and the RR 4-form (see \eqref{A2}), whereas we use the non-linear Sen action \cite{Sen:2004nf} for the flavor part 
\begin{equation}
\nn S_f = -N_c^2x_f\ell^{-3} \int\intd^5x\, V_f(\chi)\sqrt{-\det{(g_{MN}+\ell^2F_{MN}+\ell^2\pa_M\chi\pa_N\chi)}} \, , 
\end{equation}
with $F_{MN}$ the $U(1)_B$ field strength. This action describes the low-energy dynamics of a system of flavor branes and anti-branes, including the tachyonic mode, whose role is played by the di-squark scalar $\chi$ in our setup. At quadratic order in fields and derivatives, the Sen action reduces to the free kinetic terms for the scalar $\chi$, and the baryon number gauge field $A_\mu$. However, since $\chi$ is not charged under baryon number, the model can only feature a condensed phase if non-linear couplings to $A_\mu$ are included\footnote{An instability may also arise for a quadratic action if the conformal dimension of the dual operator is low enough, but only at nearly zero temperature \cite{Basu11}.}.   

The last piece of the action is a topological WZ term 
\begin{equation}
\nn S_{WZ} =  N_c^2x_f\ell^{-2}\int C_4\wedge \intd V_{WZ}(\chi) \, ,
\end{equation}
which accounts for the induced color charge carried by the scalar $\chi$. This term is directly responsible for color Higgsing, and it also plays a crucial role for the instability mechanism (as explained in section \ref{S2}). 

The parameters of the model are given by one constant $x_f$, which is proportional to the ratio of flavor to color number $N_f/N_c$, and two scalar potentials $V_f(\chi)$ and $V_{WZ}(\chi)$, that are a priori undetermined. The flavor back-reaction is fully taken into account, with $x_f$ not assumed to be small. In this work, we analyze two types of potentials, respectively associated with a second order and first order phase transition. The first type corresponds to the simplest quadratic ansatz
\begin{equation}
\nn \ell^2 V_f(\chi)-1 = C_f \chi^2 \sp \ell^2V_{WZ}(\chi) = C_{WZ} \chi^2 \, ,
\end{equation}   
with $C_f$ and $C_{WZ}$ two constant parameters, and $\ell$ the AdS length. Requiring that the operator dual to $\chi$ has conformal dimension $\D=2$ fixes the total UV mass $m_{UV}^2 = -4\ell^{-2}$, which gives a relation between $C_f$ and $C_{WZ}$. As a result, the potentials contain only one parameter, that we denote $m_{WZ}^2$ (which is proportional to $C_{WZ}$).

The second type of potentials is such that the WZ part contains an additional quartic term
\begin{equation}
\nn \ell^2 V_f(\chi)-1 = C_f \chi^2 \sp \ell^2V_{WZ}(\chi) = C_{WZ} \chi^2(1+\a \chi^2) \, .
\end{equation}
They contain two parameters, $m_{WZ}^2$ and $\a$.  

Note that our model is conformal, since it does not contain a running dilaton. This of course is an important difference with QCD, but the qualitative effect of the dilaton may not be so important in the deconfined phase.  

The equations of motion that derive from our action are listed in section \ref{S12}, and the simplest uncondensed solution is described in section \ref{S13}. The latter corresponds to the so-called DBI black brane, which is a non-linear generalization of the AdS Reissner-Nordstr\"om solution, that can be expressed analytically in terms of hypergeometric functions. 


\subsubsection*{Stability analysis}

The stability of the DBI black brane to scalar condensation is analyzed in section \ref{S2}. Our analysis uses the fact that the onset of an instability coincides with the appearance of marginally stable modes. These modes are normalizable solutions of the fluctuation equation \eqref{ms1}
\begin{equation}
\nn \d \chi''(r) - \a(r)\d\chi'(r) - \frac{\ell^2m_{eff}^2(r)}{r^2f(r)}\d\chi(r) = 0 \, ,  
\end{equation}
where $r$ is the holographic coordinate, $f(r)$ is the DBI black brane blackening function and $\a(r)$ is also known explicitly. The effective mass squared $m_{eff}^2(r)$ can be expressed as a function of a dimensionless coordinate $u$
\begin{equation}
\nn m_{eff}^2(r) = m_{eff}^2(u) = m_{UV}^2(1+u^6)^{-1}\left(1+\left(\sqrt{1+u^6}-1\right)\frac{m_{WZ}^2}{m_{UV}^2}\right) \sp  u \equiv \hat{n}^{\frac{1}{3}} r \, ,
\end{equation}
with $\hat{n}$ proportional to the quark number density\footnote{The precise relation is given in \eqref{bc7}.}. 

On general grounds, an instability is expected to occur when the effective mass squared is sufficiently negative over a large enough region in the bulk \cite{Gubser08a}. Based on the expression of $m_{eff}^2(u)$, we can therefore understand a qualitative mechanism giving rise to a condensed phase at low temperature, and negative enough $m_{WZ}^2$. 

This mechanism is illustrated in figure \ref{fms1}, and can be described as follows. For $m_{WZ}^2/m_{UV}^2>2$, the effective mass squared $m_{eff}^2(u)$ admits a minimum 
\begin{equation}
\nn m_{eff,\text{min}}^2 = \frac{m_{WZ}^4}{4(m_{WZ}^2-m_{UV}^2)} \, ,
\end{equation}
which is reached at a finite value $u_{\text{min}}$, and gets more and more negative as $m_{WZ}^2$ is decreased. On the other hand, the background DBI black brane geometry extends from $u=0$ at the boundary, to $u_H = \hat{n}^{1/3}r_H$ at the horizon. As the density $\hat{n}$ is increased (or equivalently, as the temperature is lowered), the bulk will therefore probe more negative values of $m_{eff}^2$. For sufficiently large $m_{WZ}^2/m_{UV}^2$, a critical density will exist where the instability arises. 

Going beyond these qualitative considerations, we compute in section \ref{S22} the full instability surface, by solving numerically for the marginal modes. This gives the phase diagram of the model for the case of continuous transitions, that we refer to as the ``continuous phase diagram". The result shows the expected behavior, with a phase transition ocurring at high density for sufficiently negative $m_{WZ}^2$. The dependence of the critical density on the additional parameter $x_f$ is found to be mild. 

\subsubsection*{Properties of the color superconductor}

The last section of this paper discusses several properties of the paired phase, corresponding to the holographic color superconductor. This analysis confirms explicitly that the solution realizes a phase with finite chiral condensate and Higgsing fraction, that both increase with density as shown in figure \ref{fh2} and \ref{fh2b}. 

In particular, the behavior of the Higgsing fraction is different from QCD with $N_c=3$, where the number of Higgsed generators is a finite constant below the transition temperature\footnote{The Higgsing fraction may also increase in steps as the temperature is lowered if quark masses are taken into account \cite{Alford:2001dt}.}. It is an interesting question whether this kind of behavior can be observed in large $N_c$ field theories. 

{The properties of the Higgsed phase are discussed separately for the two types of potentials mentioned above, associated respectively with a second order and first order phase transition. For the first type of potentials}, we analyze the dependence of the CSC solution on all model parameters. The most interesting result is the self-tuning of the IR effective number of colors $N_{c,IR}$ to a positive value, for any values of the parameters. This number is actually always larger than about 10\% of $N_c$, which means that these CSC solutions are never fully Higgsed. 

{The analysis of the second type of potentials indicates that first order transitions tend to generate larger condensates and Higgsing fraction. In particular, we find that more than $96\%$ of the gauge group can be Higgsed at zero temperature in this case.}

\subsection{Outlook}
\label{I4}

Our construction is the first example of a holographic color superconductor, where a paired phase is dynamically generated at finite quark number density, with a large fraction of the color group being Higgsed. As such, it opens several directions for further studies and improvements. Below is a list of example problems that would be interesting to address
\begin{itemize}
\item The thermodynamic properties of the CSC phase that we constructed are very similar to the uncondensed phase. However, it is known that pairing typically has much more impact on the transport properties \cite{Alford:2007xm}, due to the emergence of massless Goldstone modes in the spectrum. The analysis of transport in such a phase may provide interesting insight into the possible properties of dense QCD matter. Note that a setup that features standard chiral symmetry breaking in the deconfined phase (e.g. as in \cite{VQCDmu}) would a priori have similar transport properties.   
\item A related problem is to analyze the spectrum of the CSC phase. This analysis would be most interesting at finite squark mass, for which the mass dependence of the low lying spectrum can be analyzed. 
\item Another natural extension of our model is to include the dynamics of the dilaton, thus describing a non-conformal theory. The resulting action would take a similar form as in the V-QCD model \cite{VQCDvac}.
\item In \cite{Chen:2009kx}, another mechanism was considered for the generation of a CSC phase, that involved the back-reaction of the flavor fields on the NSNS and RR 2-forms. This effect may modify significantly the physics of the CSC in our setup, since the flavor back-reaction is fully taken into account. The 2-forms are usually ignored in bottom-up holography, since the background color flux makes them massive \cite{Kiritsis:2009hu}. However, for strong Higgsing, these fields become light in the IR, and may play an important role. 
\item It would also be interesting for comparison to find top-down configurations that resemble our solution in some way (in the probe limit $N_f\ll N_c$). For example, one could consider calculating an instanton solution at finite quark number density in the D3/D7 model (used e.g. in \cite{Apreda:2005yz,Chen:2009kx,Faedo:2018fjw}), by including the quartic term in the expansion of the flavor DBI action. This is likely to be harder than the problem of \cite{Faedo:2018fjw} (it may not be supersymmetric), but it is probably possible to find a numerical solution. 
\item An important new ingredient of this work is the non-trivial dynamics of the RR 4-form $C_4$, sourced by the color D3 branes. In principle, the dynamics of $C_4$ may play a role in many bottom-up constructions, even with a standard Wess-Zumino term. The latter indeed contains the magnetic coupling between $C_4$ and the Chern-Simons term, which implies in particular that baryons carry magnetic D3 charge \cite{Chen:2009kx}. It would be interesting to analyze the resulting interplay between Higgsing and holographic baryons. 
\end{itemize}

\section{The holographic model}
\label{S1}

In this section, we describe the holographic model that is analyzed in this work, which features a Higgsed phase. We first explain how the bulk action is constructed, and then present the corresponding equations of motion. The simplest uncondensed bulk solution is discussed in the last subsection. 

\subsection{Action}
\label{S11}

The model is constructed in a bottom-up approach, meaning that the bulk theory is five-dimensional, and its field content is determined from the holographic dictionary. 

We want to describe a holographic theory which exhibits a CSC phase at high baryonic density, where the di-squark operator $\phi_R^\dagger\phi_L$ condenses (with $\phi_{L/R}$ the left/right-handed squark field), and the gauge group is Higgsed. More precisely, we consider a configuration where chiral symmetry is completely broken (as in a CFL phase), and the di-squark condensate is isotropic in isospin\footnote{The possibility of a partial breaking of chiral symmetry is considered in appendix \ref{AppD}, where it is shown {that partially broken solutions can be relevant depending on model parameters. For better clarity, we present here the simpler, fully broken ansatz, leaving the general case to appendix \ref{AppD}.} }
\begin{equation}
\label{A0} \left<\phi_R^\dagger\phi_L\right> = \sigma \mathbb{I}_{N_f} \, .
\end{equation} 

To account for the above features, the bulk theory should contain a scalar $\chi$ dual to the condensate $\s$, coupled to a $U(1)_B$ baryon number gauge field $A_\mu$ and gravity, together with the RR 4-form $C_4$. The latter is always present in the bulk, since it is sourced by the color branes, but usually non-dynamical in bottom-up settings. Here its dynamics is taken into account, and describes the color group Higgsing. 

For the bulk action, we use an ansatz qualitatively similar to low energy actions arising in string theory\footnote{This approach could be called ``semi-bottom-up", with pure bottom-up corresponding to usual effective field theory. The relevance of this approach is supported by the successes of the V-QCD model \cite{VQCDvac,VQCDT,VQCDmu,Jarvinen:2021jbd}.}, which is the sum of three terms
\begin{equation}
\label{A1} S = S_c + S_f + S_{WZ} \, .
\end{equation}
The color part of the action $S_c$ is simply given by Einstein-Hilbert with a cosmological constant $\L$, together with the 4-form kinetic term
\begin{equation}
\label{A2}  S_c = N_c^2\ell^{-3}\int \intd^5x \sqrt{-g} \left(R-\L-\frac{1}{2\cdot 5!}F_5^2\right) \, ,
\end{equation}
where $F_5 = \intd C_4$, $N_c$ is the number of colors and $\ell$ the AdS length. The overall normalization of the action is in principle a free parameter, but it will not play any role in our calculation, so we fix it as in \eqref{A2}. 

For the flavor fields $A_\mu$ and $\chi$, the simplest choice would also be given by an action quadratic in derivatives. However, since the di-squark operator is not charged under baryon number, such a model does not feature a condensed phase\footnote{Except at nearly zero temperature, where condensation is triggered by the instability of the near-horizon $AdS_2$ geometry \cite{Basu11}.}. For a CSC phase to arise, the flavor action $S_f$ should therefore contain non-linear couplings between the scalar $\chi$ and the $U(1)_B$ field strength $F_{\m\n}$, for which many different forms may a priori be considered. In our model, the specific form of $S_f$ is determined by using guidance from string theory.  

In string theory, the flavor sector arises from the low energy excitations of open strings propagating on a stack of flavor branes \cite{Karch:2002sh,Sakai:2003wu}, whose dynamics is typically controlled by an action of the DBI type. Here, we will have in mind a configuration with a stack of $N_f$ pairs of space-filling branes and anti-branes\footnote{Note that the resulting configuration would have both $D3$ and $D4$ branes, which cannot happen in theories like type II string theory. {This indicates that the hypothetical five-dimensional (non-critical) string theory dual to QCD is more similar to type 0 string theory \cite{Polyakov:1998ju,Kiritsis:2009hu},}{where odd and even branes can coexist. This kind of D3-D4 setup may also arise from the dimensional reduction of type IIB on the internal 5-sphere, as in the D3-D7 model discussed in appendix \ref{AppA1}.}} $D4-\bar{D4}$, with $N_f$ the number of flavors. Having both branes and anti-branes ensures that the boundary theory has a chiral isospin symmetry $U(N_f)_L\times U(N_f)_R$, as in massless QCD or ($\mathcal{N}=1$) superQCD (SQCD). The left-handed factor arises from the gauge theory on the $D4$ branes, whereas the right-handed one arises from the anti-branes. 

In this brane setting, $A_\mu$ is the gauge field for the vector part of the abelian subgroup $U(1)_{L+R}$. The scalar $\chi$ arises from the lowest excitation of open strings connecting branes to anti-branes. As such, its condensation also breaks chiral symmetry, as expected for a field dual to the di-squark $\phi_R^\dagger\phi_L$. 

In string theory on a  flat background, $\chi$ would be a tachyonic mode, and its coupled dynamics with the worldvolume gauge fields would be controlled by the Sen action \cite{Sen:2004nf,Kutasov:2000qp,Minahan:2000tf,Garousi:2000tr,Takayanagi:2000rz}. Here, we will choose the flavor action to be of a similar form, with an expression given by
\begin{equation}
\label{A3} S_f = -N_c^2x_f\ell^{-3} \int\intd^5x\, V_f(\chi)\sqrt{-\det{(g_{MN}+\ell^2F_{MN}+\ell^2\pa_M\chi\pa_N\chi)}} \, , 
\end{equation}
where $\ell$ is the AdS length, and $x_f\equiv N_f/N_c$. For the case of a flat background, string field theory predicts a tachyon potential of the form \cite{Takayanagi:2000rz}
\begin{equation}
\label{A4} V_{f,\text{SFT}}(\chi) = \ex^{-a\chi^2} \sp a>0 \, .
\end{equation}
In our setup, we allow for a general $V_f(\chi)$, which may contain several parameters. 

The action \eqref{A3} has been considered in other holographic setups, and shown to give a consistent description for the dynamics of chiral symmetry breaking \cite{Casero:2007ae,VQCDvac,VQCDT,VQCDmu}. In those works the tachyon field $\chi$ was interpreted as the dual of the quark bilinear $\bar{\psi}_R\psi_L$, whereas here we see it as being dual to the di-squark, as in top-down examples \cite{Chen:2009kx,Faedo:2018fjw}. There is no important distinction between the two cases however, and the tachyon field could equally well be considered as the dual of $\bar{\psi}_R\psi_L$, upon appropriately modifying its UV mass (see below in \eqref{ms6}). 

The branes action also contains a Wess-Zumino (WZ) term, which couples the flavor fields to the color RR forms. Here, it gives a coupling between the 4-form $C_4$ and the flavor scalar $\chi$, with RR gauge-invariance requiring a term of the form 
\begin{equation}
\label{A5} S_{WZ} =  N_c^2x_f\ell^{-2}\int C_4\wedge \intd V_{WZ}(\chi) \, ,
\end{equation}
where $V_{WZ}$ is an additional, a priori unfixed, potential. The WZ term \eqref{A5} plays an essential role for the CSC dynamics, since it implies that the condensation of $\chi$ induces color brane charge on the flavor branes, which corresponds to Higgsing of the color group.  

Note that a term like \eqref{A5} does not appear in the  flat space WZ term derived from string field theory \cite{Takayanagi:2000rz,Casero:2007ae}. We find it likely however that the generalization to a string theory that contains both even and odd RR forms may contain a term of this type. Alternatively, this kind of term can certainly arise from the dimensional reduction of a (critical) string-theoretic setting, as shown by the example in appendix \ref{AppA1}. 

Equations \eqref{A2}, \eqref{A3} and \eqref{A5} together define the action of our holographic model. Its parameters are given by one constant $x_f$, and two functions $V_f$ and $V_{WZ}$. Note that there is some degeneracy between $x_f$ and the overall normalization of the potentials. In the following, we use this degeneracy to fix the value of $V_f(0)$ as
\begin{equation}
\label{A5b} \ell^2 V_f(0) = 1 \, .
\end{equation}
The ratio of flavor to color number $x_f$ is not assumed to be small, so that the back-reaction of the flavor sector is fully taken into account.

At this point, let us give some comments on the field theory dual of our model \eqref{A1}. As explained before, we have in mind describing the dynamics of a theory relatively close to QCD (or SQCD). QCD and SQCD are not conformal theories, so that the color sector of the gravitational dual in principle contains a running dilaton, which is not included in our model. The generalization of \eqref{A3} that includes the dilaton is the V-QCD action \cite{VQCDvac}. The dilaton is very important to properly describe the IR regime, where the theory becomes strongly-coupled. However, we do not expect an important qualitative difference in the deconfined phase that interests us, where the dilaton running is mild.

\subsection{Equations of motion}
\label{S12}

The general equations of motion associated with the action \eqref{A1} are written in appendix \ref{AppA}. In this work, we are interested in black brane solutions corresponding to homogeneous quark matter at finite temperature and density, in which a chiral condensate may form. The appropriate ansatz for the bulk fields is given by
\begin{equation}
\label{E1} \intd s^2 = \ex^{2A(r)} \left(-f(r)\intd t^2 + f(r)^{-1}\intd r^2 + \intd\vec{x}^2\right) \, ,
\end{equation} 
\begin{equation}
\label{E2} A \equiv A_\m\intd x^\m = A_t(r) \intd t \sp \chi = \chi(r) \, ,
\end{equation}
\begin{align}
\label{E3} F_5= \sqrt{-g} \frac{w(r)}{\ell}  \intd t\wedge\intd r\wedge\intd x^1\wedge\intd x^2\wedge\intd x^3 \, ,
\end{align}
where $r$ is the holographic coordinate. It is defined such that the boundary lies at $r=0$, and the horizon at $r=r_H$, where $f(r_H)=0$. Note the form of the ansatz for the color flux $F_5$, with $w$ a function of $r$. In absence of Higgsing, $w$ would be a constant $w_0$ \cite{Kiritsis:2009hu}, whereas Higgsing translates into a decrease of $w$ from the UV to the IR.  

Substituting the ansatz \eqref{E1}-\eqref{E3} into the equations of motion \eqref{eom1}-\eqref{eom4} gives the system of equations obeyed by the ansatz fields 
\begin{equation}
\label{E4} \pa_r\left(\ex^A\Xi^{-\frac{1}{2}}V_f(\chi)A_t'\right) = 0 \, ,
\end{equation}
\begin{equation}
\label{E5} w' = x_f\ell^2\pa_rV_{WZ}(\chi) \, ,
\end{equation}
\begin{equation}
\label{E6} \ex^{-5A}\pa_r\left(\ex^{3A}f\chi'\right) + \pa_r\log{\left(\Xi^{-\frac{1}{2}}V_f(\chi)\right)}\ex^{-2A}f \chi' - \Xi \left(\frac{V_f'(\chi)}{V_f(\chi)}+ \Xi^{-\frac{1}{2}}\frac{V_{WZ}'(\chi)}{V_f(\chi)}w\right) = 0 \, ,
\end{equation}
\begin{equation}
\label{E7} A'' - (A')^2 + \frac{x_f}{6}\Xi^{-\frac{1}{2}}V_f(\chi)(\chi')^2 = 0 \, ,
\end{equation}
\begin{equation}
\label{E8} A'f' + 4f(A')^2 + \frac{1}{6}\ex^{2A}\left(2\L+2x_fV_f(\chi)\Xi^{-\frac{1}{2}}+\frac{w^2}{\ell^2}\right) \, ,
\end{equation}
where we defined
\begin{equation}
\label{E9} \Xi(r) \equiv 1 + \ex^{-2A(r)}f(r)\ell^2\chi'(r)^2-\ex^{-4A(r)}\ell^4A_t'(r)^2 \, .
\end{equation}
The first two equations for $A_t(r)$ and $w(r)$ integrate trivially to
\begin{equation}
\label{E10} w(r) =  w_0 +x_f\ell^2(V_{WZ}(\chi(r))-V_{WZ}(0)) \, ,
\end{equation}
\begin{equation}
\label{E11} A_t'(r) = \ex^{-A(r)}\Xi(r)^{\frac{1}{2}}\frac{\hat{n}}{\ell V_f(\chi(r))}\,,
\end{equation}
with $w_0$ and $\hat{n}$ two integration constants. $w_0$ is a parameter of the model, related to the color brane tension \cite{Kiritsis:2009hu}, whereas $\hat{n}$ is proportional to the quark number density $n_q$, with the precise relation given below in \eqref{bc7}. 

Note that \eqref{E11} gives an algebraic equation for $A_t'(r)$, which can be solved to express it in terms of the other fields as 
\begin{equation}
\label{E13} A_t'(r) = -\frac{\hat{n}}{\ex^A \ell V_f(\chi)}\sqrt{\frac{1+\ex^{-2A}f\ell^2(\chi')^2}{1+\ex^{-6A}\ell^2\hat{n}^2V_f(\chi)^{-2}}} \, .
\end{equation}

\subsubsection{Boundary conditions}
\label{S121}

To fully specify the black brane solutions, the equations of motion \eqref{E4}-\eqref{E8} should be supplemented with appropriate boundary conditions. These include both UV conditions, with a choice of boundary sources, and IR conditions at the horizon, where regularity is imposed. Note that $w$ obeys a first order equation \eqref{E5}, so it is not necessary to discuss it separately.

At finite temperature, the regular behavior of the fields near the horizon $(r\to r_H)$ is given by :
\begin{equation}
\label{bc1} A(r) = A(r_H) + \OO(r-r_H) \sp f(r) = f'(r_H)(r-r_H)+\OO(r-r_H)^2 \, ,
\end{equation}
\begin{equation}
\label{bc2} A_t(r) = A_t'(r_H)(r-r_H) + \OO(r-r_H)  \sp \chi(r)=\chi(r_H)+\OO(r-r_H) \, .
\end{equation}
$A(r_H),f'(r_H),A_t'(r_H),\chi(r_H)$ are the IR parameters of the solution, that fully determine the higher order coefficients. $A(r_H)$ and $f'(r_H)$ correspond to the residual freedom of coordinates rescaling. They are fixed by setting $f$ to 1 at the boundary, and choosing the unit in which dimensionful boundary observables are measured. A possible choice is to measure things in units of the temperature, which amounts to setting $T=-f'(r_H)/(4\pi)$ to 1. 

As for $\chi(r_H)$, it controls the boundary source for the dual operator $\phi^\dagger_R\phi_L$. Since we are interested in spontaneous symmetry breaking, $\chi(r_H)$ is fixed such that the source vanishes.

The remaining coefficient $A_t'(r_H)$ controls the only actual parameter of the solution (in addition to model parameters), that is $\mu/T$, with $\mu$ the quark number chemical potential. The latter appears at leading order in the near-boundary ($r\to 0$) expansion of the gauge field 
\begin{equation}
\label{bc3} A_t(r) = \mu - A_t^{(2)} r^2\left(1+\OO(r^2)\right) \, ,
\end{equation} 
whereas the UV behavior of the other ansatz fields is given by
\begin{equation}
\label{bc4} A(r) = -\log\left(\frac{r}{\ell}\right) - A^{(4)}r^4\left(1+\OO(r^2)\right) \sp f(r) = 1 - f^{(4)}r^4\left(1+\OO(r^2)\right)  \, ,
\end{equation}
\begin{equation}
\label{bc5} \chi(r) = \chi^{(\D)}r^{\D}\left(1+\OO(r^2)\right)  \, ,
\end{equation}
where $\D=2$ is the conformal dimension of the di-squark operator\footnote{{Note that this corresponds to the special case $\D = d/2$ - with $d$ the boundary space-time dimension - for which the UV fall-off of the bulk scalar field is of the form $\chi(r) \underset{r\to 0}{=} \chi_- r^\D \log(r/r_0) + \chi_+ r^\D + \mathcal{O}(r^{\D+2})$. Even though the two independent near-boundary behaviors are both normalizable in this case, the only quantization compatible with conformal invariance is to treat the coefficient of the log term as the source \cite{Klebanov:1999tb}.}{ Setting the source to zero therefore implies the behavior in \eqref{bc5}.}}. The coefficients of \eqref{bc3}-\eqref{bc5} are related to the expectation values of the dual operators\footnote{These relations can be established in a standard way from variations of the renormalized on-shell action \cite{Klebanov:1999tb,deHaro:2000vlm,Skenderis:2002wp}. {Note that, since the boundary dimension $d=4$ is even, there is a possible matter conformal anomaly in the system. However, this anomaly cancels when the source of the scalar operator is set to zero, as stated in the second equation of \eqref{bc6}.}} as
\begin{equation}
\label{bc6} n_q = 2N_c^2x_f \ell^2V_f(0) A_t^{(2)} \sp p = \frac{1}{3}\EE = N_c^2f^{(4)} \sp \left<\phi_R^\dagger\phi_L\right> = \chi^{(\D)} \, ,
\end{equation}
where $n_q$ is the quark number density, $p$ the pressure and $\EE$ the energy density. In particular, from the first relation in \eqref{bc6}, the integration constant $\hat{n}$ in \eqref{E11} is seen to be related to $n_q$ by
\begin{equation}
\label{bc7} n_q = N_c^2 x_f \hat{n} \, .
\end{equation}


\subsection{The uncondensed solution}
\label{S13}

The solutions of the bulk equations of motion \eqref{E4}-\eqref{E8} correspond to different equilibrium states of the boundary theory. In our setup, there exist two kinds of solutions, depending on whether a scalar condensate forms or not. To determine whether the condensed phase dominates in some regime, it is necessary to compute both solutions, and compare their free energies. In this subsection, we begin with a description of the uncondensed solution, while the condensed phase is addressed in the following sections. 

The uncondensed solution exists for any temperature, and corresponds to the DBI black brane geometry \cite{VQCDmu}, which is a non-linear generalization of AdS Reissner-Nordstr\"om. When $\chi=0$, $w=w_0$ is a constant, and the equations of motion for the metric and gauge field \eqref{E7},\eqref{E8},\eqref{E11} become
\begin{equation}
\label{u1} A_t' = \ex^{-A}G^{\frac{1}{2}}\ell\hat{n}\, ,
\end{equation}
\begin{equation}
\label{u2} A'' - (A')^2 = 0 \, ,
\end{equation}
\begin{equation}
\label{u3} A'f' + 4f(A')^2 + \frac{1}{6}\ex^{2A}\left(2\L+\frac{2x_f}{\ell^2}G^{-\frac{1}{2}}+\frac{w_0^2}{\ell^2}\right) \, ,
\end{equation}
where $G(r)$ is defined as
\begin{equation}
\label{u4} G(r) \equiv 1 -\ex^{-4A(r)}\ell^4A_t'(r)^2 \, ,
\end{equation}
and we recall that we fixed $\ell^2V_f(0)=1$. 

The equation for $A(r)$ \eqref{u2} is the same as in AdS, which is solved by
\begin{equation}
\label{u5} A(r) = -\log{\left(\frac{r}{\ell}\right)} \, ,
\end{equation}
whereas \eqref{u1} results in simple expressions for $A_t'$ and $G$
\begin{equation}
\label{u6} A_t'(r) = -\hat{n}r\left(1+\hat{n}^2r^6\right)^{-\frac{1}{2}} \sp  G(r) = \left(1+\hat{n}^2r^6\right)^{-1} \, .
\end{equation}
The near-boundary limit of \eqref{u3} then gives a relation between the cosmological constant $\L$ and the AdS length 
\begin{equation}
\label{u7} \L = -\ell^{-2}\left(12 + x_f + \frac{1}{2}w_0^2 \right)\, .
\end{equation}

Given \eqref{u6} and \eqref{u7}, \eqref{u1} and \eqref{u3} can be integrated to express $A_t$ and $f$ in terms of hypergeometric functions \cite{VQCDmu}
\begin{equation}
\label{u8} A_t(r) = \frac{1}{2}\hat{n}r_H^2\left(X(r_H)-X(r)\right) \sp X(r) \equiv \left(\frac{r}{r_H}\right)^2 {}_2F_1\left(\frac{1}{3},\frac{1}{2},\frac{4}{3};-\hat{n}^2r^6\right) \, ,
\end{equation}
\begin{equation}
\label{u9} f(r) = 1 - \left(\frac{r}{r_H}\right)^4\left(1+Y(r_H)\right) + Y(r) \!\!\sp\! Y(r) \equiv \frac{x_f}{12}\left(\!1- {}_2F_1\left(-\frac{2}{3},-\frac{1}{2},\frac{1}{3};-\hat{n}^2r^6\right)\!\right)\! ,
\end{equation}
where we recall that $r_H$ refers to the horizon radius. Equations \eqref{u5},\eqref{u8} and \eqref{u9} give the explicit form of the DBI black brane solution. 

To analyze the phase diagram of the model, we will need to know the free energy of the corresponding thermal state. For this, it is sufficient to compute the pressure $p$ and chemical potential $\mu$. The latter may be extracted from the UV behavior of the solution using the results of section \ref{S12}, which gives 
\begin{equation}
\label{u10} \mu(r_H,\hat{n}) = X(r_H) \sp p(r_H,\hat{n}) = N_c^2r_H^{-4}(1+Y(r_H)) \, ,
\end{equation} 
where the $\hat{n}$ dependence appears in the expressions for $X$ and $Y$ \eqref{u8}-\eqref{u9}. Even though $(r_H,\hat{n})$ are convenient coordinates to express the thermodynamic functions in this background, the physically meaningful parametrization of the phase diagram is rather in terms of temperature $T$ and density $\hat{n}$ (canonical ensemble), or temperature and chemical potential (grand-canonical ensemble). Expressing $p$ and $\mu$ as functions of $T$ and $\hat{n}$ requires to compute $r_H(T,\hat{n})$, by inverting the relation
\begin{equation}
\label{u11}  T = -\frac{1}{4\pi} f'(r_H) = (\pi r_H)^{-1}\left[1 + \frac{x_f}{12}\left(1-\sqrt{1+\hat{n}^2r_H^6}\right)\right] \, .
\end{equation}
This does not provide a closed form expression in general for $r_H(T,\hat{n})$, except for the limiting cases $\hat{n}=0$, and $T=0$, for which \eqref{u11} gives
\begin{equation}
\label{u12} r_H(T,\hat{n}=0) = (\pi T)^{-1} \sp r_e(\hat{n})\equiv r_H(T=0,\hat{n}) = \hat{n}^{-\frac{1}{3}}\left[\left(1+\frac{12}{x_f}\right)^2- 1\right]^{\frac{1}{6}} \, .
\end{equation}

\section{Stability analysis}
\label{S2}

In this section, we analyze the stability of the uncondensed solution presented above to the formation of a scalar hair, which would signal a transition to the condensed phase. We do this by looking for marginally stable modes as the temperature is varied.
The results allow us to construct the phase diagram for the case of continuous phase transitions, which exhibits a CSC phase at low temperature for appropriate values of model parameters. Since the underlying theory is conformal, the dependence of the phase diagram on temperature and density is via the dimensionless ratio 
\begin{equation}
\label{sa1} \mathcal{T} \equiv \frac{T}{\hat{n}^{1/3}} \, .
\end{equation}

\subsection{Marginally stable mode}
\label{S21}

At a point $\mathcal{T}_c$ where an instability develops, one of the poles of the scalar retarded two-point function $G^R(\omega)$ moves from the lower half complex plane to the upper half plane. In the gravitational dual, the poles of the retarded correlator correspond to quasi-normal modes of the background \cite{Son:2002sd}. Precisely at $\mathcal{T}_c$, the bulk solution therefore exhibits a normalizable marginally stable mode, that is with $\Im{(\omega)} = 0$. Conversely, the existence of such marginally stable modes is a good indication for the presence of an instability \cite{Gubser08a}. 

In our case, the marginally stable modes that arise are actually time independent ($\omega=0$), so we may consider a small fluctuation of the scalar field of the form $\d\chi = \d\chi(r)$. The equation obeyed by this fluctuation is then obtained by linearizing the scalar equation of motion \eqref{E6}, on top of the DBI black brane background. This gives\footnote{Note that we imposed $V_f'(0)=V_{WZ}'(0)=0$. Requiring that the bulk admits AdS asymptotics imposes one condition $V_f'(0)+w_0V_{WZ}'(0)=0$. The additional condition ensures that a hairless solution exists for any temperature.}
\begin{equation}
\label{ms1} \d\chi''(r) -\left(\frac{3}{r}-\frac{f'(r)}{f(r)}+\frac{G'(r)}{2G(r)}\right)\d\chi'(r) - \frac{\ell^2m_{eff}^2(r)}{r^2f(r)}\d \chi(r) = 0 \, , 
\end{equation}
where 
\begin{equation}
\label{ms2} G(r) = \left(1+\hat{n}^2r^6\right)^{-1} \, ,
\end{equation}
and the expression for $f(r)$ can be found in \eqref{u9}. 

The effective mass squared is given by
\begin{equation}
\label{ms3} m_{eff}^2(r) \equiv G(r)\left(m_f^2+G(r)^{-\frac{1}{2}}m_{WZ}^2\right) \, ,
\end{equation}
where we defined 
\begin{equation}
\label{ms4} m_f^2 \equiv V_f''(0) \sp m_{WZ}^2 \equiv w_0 V_{WZ}''(0) \, .
\end{equation}
Near the boundary, where $G(r)$ goes to 1, $m_{\text{eff}}^2$ should go to the UV mass of the scalar. This implies a relation between the two mass parameters
\begin{equation}
\label{ms5} m_f^2 + m_{WZ}^2 = m_{UV}^2 \, ,
\end{equation}
where $m_{UV}$ is fixed in terms of the di-squark conformal dimension
\begin{equation}
\label{ms6} m_{UV}^2 = \D(\D-4)\ell^{-2} = -4\ell^{-2} \, .
\end{equation}
Equation \eqref{ms5} then allows to write the effective mass \eqref{ms3} in terms of a single parameter $m_{WZ}^2$
\begin{equation}
\label{ms7} m_{eff}^2(r) = m_{UV}^2G(r)\left(1 + \left(G(r)^{-\frac{1}{2}}-1\right)\frac{m_{WZ}^2}{m_{UV}^2}\right) \, .
\end{equation}

It is expected that the scalar will condense for an effective mass squared that is sufficiently negative over a sufficiently large region of the bulk \cite{Gubser08a}. The qualitative properties of $m_{eff}^2(r)$ for various values of $m_{WZ}^2$ and $\mathcal{T} \equiv T/\hat{n}^{1/3}$ can therefore give good indications on where we expect an instability to develop. In particular, we can readily make the following two observations: 
\begin{itemize}
\item $G(r)$ in \eqref{ms2} is smaller than 1, so the effective mass \eqref{ms7} can only decrease in the bulk for $m_{WZ}^2 < 0$. Interestingly, this is the same condition that one gets from requiring that the effective number of color $w(r)$ can only decrease away from the boundary. Indeed, the near-boundary limit of \eqref{E10} is given by
\begin{equation}
\label{ms8} w(r) = w_0 +\frac{x_f}{2w_0}(m_{WZ}\ell)^2\chi(r)^2 + \OO(\chi^3) \, .
\end{equation}
\item In the limit of vanishing baryon density ($\mathcal{T}\to\infty$), $G(r)$ goes to 1, and the effective mass \eqref{ms7} is a constant. Condensation can therefore only occur at finite baryon density. 
\end{itemize}

\begin{figure}[h!]
\begin{center}
\includegraphics[scale=1.3]{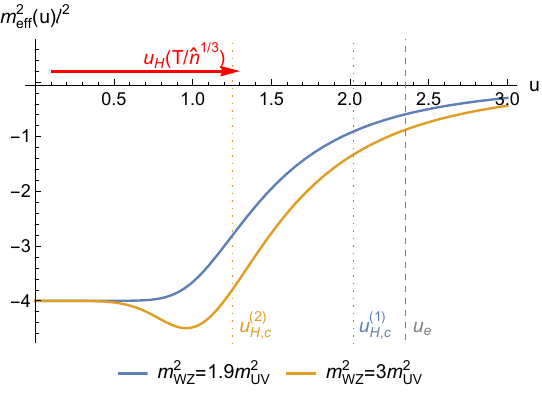}
\caption{$m_{eff}^2(u)$ for two values of $m_{WZ}^2/m_{UV}^2$: 1.9 (blue) and 3 (orange). The coordinate $u=\hat{n}^{1/3}r$ extends from 0 at the boundary, to a finite value $u_H(\mathcal{T})$ at the horizon ($\mathcal{T} \equiv T/\hat{n}^{1/3}$). At zero temperature, $u_H$ reaches the extremal value $u_e$ (expressed in \eqref{ms11}), which is indicated by the gray dashed line. The qualitative mechanism for the low temperature instability at $m_{WZ}^2 < 2m_{UV}^2$ can be understood from the shape of the orange line. As the temperature is lowered, $u_H$ increases (red arrow), so that $m_{eff}^2$ is lower than $m_{UV}^2$ over a larger and larger region of the bulk. The onset of the instability for $m_{WZ}^2=3m_{UV}^2$ is denoted $u_{H,c}^{(2)}$, and indicated by the orange dotted line. Note that an instability also occurs at high density for $m_{WZ}^2=1.9m_{UV}^2$ (at $u_{H,c}^{(1)}$, shown as the blue dotted line), even though $m_{eff}^2(u)$ is a monotonically increasing function. In this case, the behavior of $m_{eff}^2$ alone cannot explain the instability. }
\label{fms1}
\end{center}
\end{figure}

To investigate in more details the properties of $m_{eff}^2(r)$, it is useful to note from \eqref{ms2} that it is actually a function of the dimensionless coordinate 
\begin{equation}
\label{ms9} u \equiv \hat{n}^{\frac{1}{3}} r \, ,
\end{equation}
so that the density dependence reduces to a rescaling of $r$. As a function of $u$, $m_{eff}^2$ goes from $m_{UV}^2$ at $u=0$, to $0$ at $u=\infty$, and its main feature is to possess a minimum, whose location and value are given by 
\begin{equation}
\label{ms10} u_{\text{min}} =
\left\{
\begin{array}{ll}
	\left[4\left(1-\frac{m_{UV}^2}{m_{WZ}^2}\right)^2-1\right]^{\frac{1}{6}} & \, ,\,\, m_{WZ}^2/m_{UV}^2 \geq 2 \\
	\, 0 & \, ,\,\, m_{WZ}^2/m_{UV}^2 \leq 2 
\end{array}
\right. \, ,
\end{equation}
\begin{equation}
\label{ms10b} m_{eff}^2(u_{\text{min}}) =
\left\{
\begin{array}{ll}
	  \frac{m_{WZ}^4}{4(m_{WZ}^2-m_{UV}^2)} & \, ,\,\, m_{WZ}^2/m_{UV}^2  \geq 2 \\
	\, m_{UV}^2 & \, ,\,\, m_{WZ}^2/m_{UV}^2  \leq 2 
\end{array}
\right. \, .
\end{equation}
The behavior of the effective mass squared is therefore qualitatively different depending on whether $m_{WZ}^2/m_{UV}^2$ is smaller or larger than 2. This is  illustrated in figure \ref{fms1}, which shows $m_{eff}^2(u)$ for $m_{WZ}^2/m_{UV}^2 = 1.9$ and $3$.


For a fixed value of $m_{WZ}^2$, the mechanism that leads to an instability at high density can be understood pictorially as shown in figure \ref{fms1}. As stated before, the shape of $m_{eff}^2(u)$ itself does not depend on density. However, the bulk only extends over a finite range of $u$, from 0 at the boundary to $u_H\equiv \hat{n}^{1/3}r_H(T,\hat{n})$ at the horizon. $u_H$ increases with density, up to an $x_f$-dependent constant 
\begin{equation}
\label{ms11} u_e \equiv \hat{n}^{\frac{1}{3}} r_e(\hat{n}) = \left[\left(1+\frac{12}{x_f}\right)^2- 1\right]^{\frac{1}{6}} \, ,
\end{equation}
which is computed from \eqref{u12}. Now, combining \eqref{ms10} and \eqref{ms11}, we see that, if $m_{WZ}^2$ is negative enough, and $x_f$ small enough\footnote{Equation \eqref{ms11} actually suggests a very mild dependence on $x_f$ for $x_f\gg 1$.}, then a condensed phase will form at high density. The precise density beyond which the instability sets in is computed numerically in the next subsection. 


\subsection{Continuous phase diagram}
\label{S22}

In the previous section, we described the qualitative mechanism underlying the low temperature instability in our model. We will now proceed to a quantitative analysis of the instability, by computing numerically the marginal modes that solve \eqref{ms1}. {When the phase transition is continuous - as for the potentials of section \ref{S31}}{ - these results give access to the phase diagram of the model. For first order transitions as in section \ref{S32},}{ the instability analysis below still gives a useful characterization of the phase structure, but the calculation of the full phase diagram requires more work. Following the previous discussion, we will refer to the instability surface as the ``continuous phase diagram".}

\begin{figure}[h!]
\begin{center}
\includegraphics[scale=1.3]{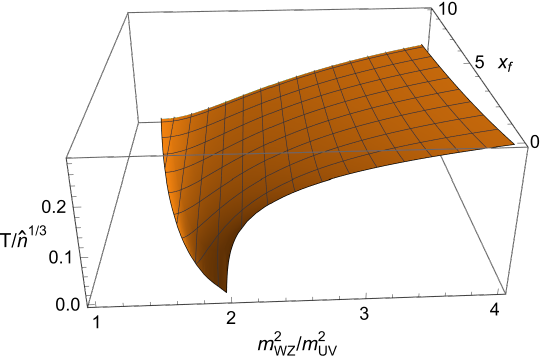}
\caption{Phase diagram of our model in the three-dimensional parameter space $(m_{WZ}^2/m_{UV}^2,x_f,T/\hat{n}^{1/3})$, for (any) choices of potentials such that the phase transition is continuous. The paired phase is located below the orange surface, where the phase transition happens.}
\label{fpd1}
\end{center}
\end{figure} 

The continuous phase diagram is shown in figure \ref{fpd1}. As should be clear from section \ref{S21}, it does not depend on the details of the scalar potentials, but only on three parameters: the model parameters $m_{WZ}^2/m_{UV}^2$ and $x_f$, and temperature in units of density $\mathcal{T} \equiv T/\hat{n}^{1/3}$. The general structure is as anticipated, with a condensed phase existing at high density for large enough $m_{WZ}^2$. 

\begin{figure}[h!]
\begin{center}
\includegraphics[scale=0.8]{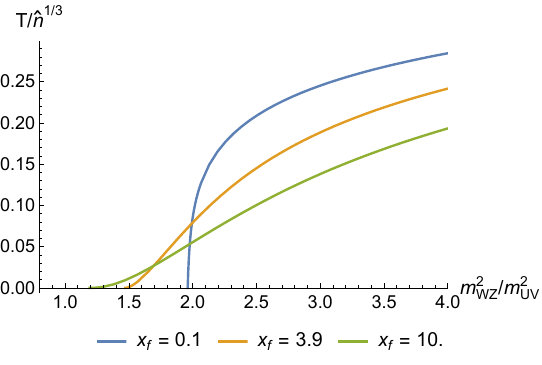}\hspace{0.5cm}
\includegraphics[scale=0.75]{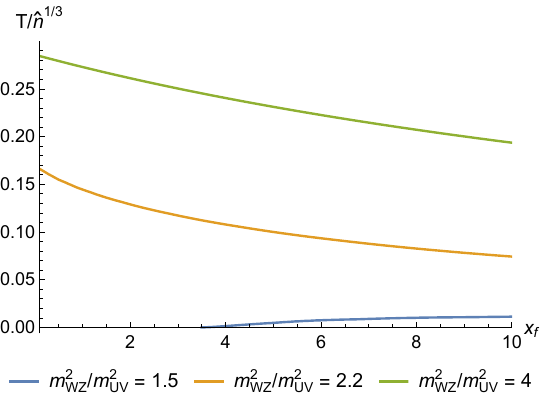}
\caption{Cuts of the continuous phase diagram (figure \ref{fpd1}) at constant $x_f$ (Left) and constant $m_{WZ}^2$ (Right).}
\label{fpd2}
\end{center}
\end{figure} 

\begin{figure}[h!]
\begin{center}
\includegraphics[scale=1]{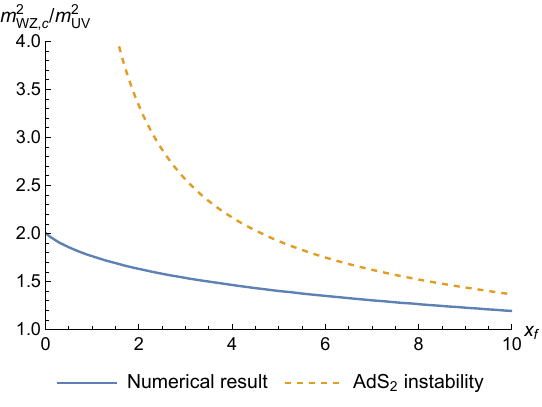}
\caption{The continuous phase transition line at zero temperature, with the paired phase located above the blue line. The orange dashed line indicates the onset of the $AdS_2$ instability, as determined by \eqref{pd4}.}
\label{fpd3}
\end{center}
\end{figure}

Figure \ref{fpd2} provides a more detailed view at fixed $m_{WZ}^2$ and $x_f$. For a given value of $x_f$, the condensed phase is seen to exist for $m_{WZ}^2$ above a certain value $m_{WZ,c}^2$, which depends on $x_f$ as shown in figure \ref{fpd3}. Above this critical value, the transition temperature $\mathcal{T}_c$ increases monotonically with $m_{WZ}^2$. At large $m_{WZ}^2/m_{UV}^2$, a simple argument presented in Appendix \ref{AppB} shows that it behaves as 
\begin{equation}
\label{pd1} \mathcal{T}_c\, \underset{m_{WZ}^2/m_{UV}^2\gg 1}{\sim} \left(\frac{m_{WZ}^2}{m_{UV}^2}\right)^{\frac{1}{6}} \, , 
\end{equation}
with an order 1 constant coefficient. 

Regarding the $x_f$ dependence of the continuous phase diagram, figures \ref{fpd1} and \ref{fpd2} show the existence of two different regimes depending on the value of $m_{WZ}^2$. For $m_{WZ}^2/m_{UV}^2>2$, the dependence is rather mild, with the critical temperature slowly decreasing with $x_f$. In contrast, for $m_{WZ}^2/m_{UV}^2<2$, $x_f$ plays a more important role, which is mainly to induce a decrease of the critical mass $m_{WZ,c}^2$ as shown in figure \ref{fpd3}.

We shall conclude our discussion of the continuous phase diagram by commenting on the zero temperature limit. At $T=0$, the uncondensed background is an extremal black brane, which contains an $AdS_2$ near-horizon geometry. In this case, there is a well-known mechanism for an instability to develop, which occurs when the effective mass at the horizon violates the $AdS_2$ Breitenlohner-Freedman (BF) bound \cite{Jensen:2010ga,Iqbal:2010eh}
\begin{equation}
\label{pd2} m_{eff}^2(u_e) < -\frac{1}{4\ell_2^2} \, ,
\end{equation}
with $\ell_2$ the $AdS_2$ length. $\ell_2$ and $m_{eff}^2(u_e)$ are given by\footnote{See appendix \ref{AppB} for a derivation of $\ell_2$.} 
\begin{equation}
\label{pd3} \frac{\ell_2}{\ell} = \frac{1}{2\sqrt{3}}\sqrt{\frac{12+x_f}{6+x_f}} \sp m_{eff}^2(u_e) = m_{UV}^2 \left(1+\frac{12}{x_f}\right)^2 \left[1+\frac{12}{x_f}\frac{m_{WZ}^2}{m_{UV}^2}\right] \, ,
\end{equation} 
where we used the expression \eqref{u12} for the extremal horizon radius. Combining \eqref{pd2} and \eqref{pd3} implies that the near-horizon $AdS_2$ becomes unstable when 
\begin{equation}
\label{pd4} m_{WZ}^2 < m_{WZ,2}^2(x_f) \equiv m_{UV}^2\left(\frac{9}{8}-\frac{x_f}{48}+\frac{9}{2 x_f} \right) \, .
\end{equation}

The line corresponding to \eqref{pd3} is shown in figure \ref{fpd3}, together with the numerically computed transition line at zero temperature $m_{WZ,c}^2(x_f)$. For the range of $x_f$ that is shown, the phase transition is observed to always happen before the $AdS_2$ instability turns on. However, figure \ref{fpd3} also indicates that the two lines tend to get closer as $x_f$ is increased, meaning that the instability may become of the $AdS_2$ type at large $x_f$. We checked that it does happen for $x_f \gtrsim 20$, where the zero-temperature transition becomes of the Berezinskii-Kosterlitz-Thouless (BKT) class, as generally expected for this kind of instability \cite{Jensen:2010ga,Iqbal:2010eh}.

\section{Properties of the Higgsed phase}
\label{S3}

We have seen that our model exhibits a condensed phase in the high density regime. In this section, we will discuss the properties of this phase, related in particular to the gauge group Higgsing. Unlike marginally stable modes of the DBI black brane, it is clear from the equations of motion \eqref{E4}-\eqref{E8} that the condensed solution depends on the specific choice of flavor potentials $V_f(\chi)$ and $V_{WZ}(\chi)$. {In the following, we analyze two types of potentials, respectively associated with a second order and a first order phase transition. }

\subsection{Quadratic potentials}

\label{S31}

We start by analyzing the simplest case where the potentials are pure mass terms
\begin{equation}
\label{h1} \ell^2V_f(\chi) = 1+\frac{1}{2}m_f^2\ell^2 \chi^2 \sp \ell^2V_{WZ}(\chi) = \frac{1}{2w_0}m_{WZ}^2\ell^2 \chi^2 \, , 
\end{equation} 
where we used the notations \eqref{ms4}. As noted in section \ref{S1}, string theory rather predicts potentials of an exponential form \eqref{A4}, which would suggest to replace \eqref{h1} by 
\begin{equation}
\nn \ell^2V_{f,e}(\chi) = \ex^{a\, m_{UV}^2\ell^2 \chi^2}\left(1+\frac{1}{2}(m_f^2-2am_{UV}^2)\ell^2\chi^2\right) \sp a>0 \, , 
\end{equation} 
\begin{equation}
\label{h2} \ell^2V_{WZ,e}(\chi) = \ex^{a\, m_{UV}^2\ell^2 \chi^2}\left(1+\frac{1}{2}(w_0^{-1}m_{WZ}^2-2am_{UV}^2)\ell^2\chi^2\right) \, .
\end{equation}
We checked however that using \eqref{h2} instead of \eqref{h1} does not induce important qualitative differences. Essentially, this is because the value of the scalar $\chi$ always remains relatively small, for both types of potentials. In the following, we choose to work with the simpler ansatz \eqref{h1}.   

\begin{figure}[h!]
\begin{center}
\includegraphics[scale=0.55]{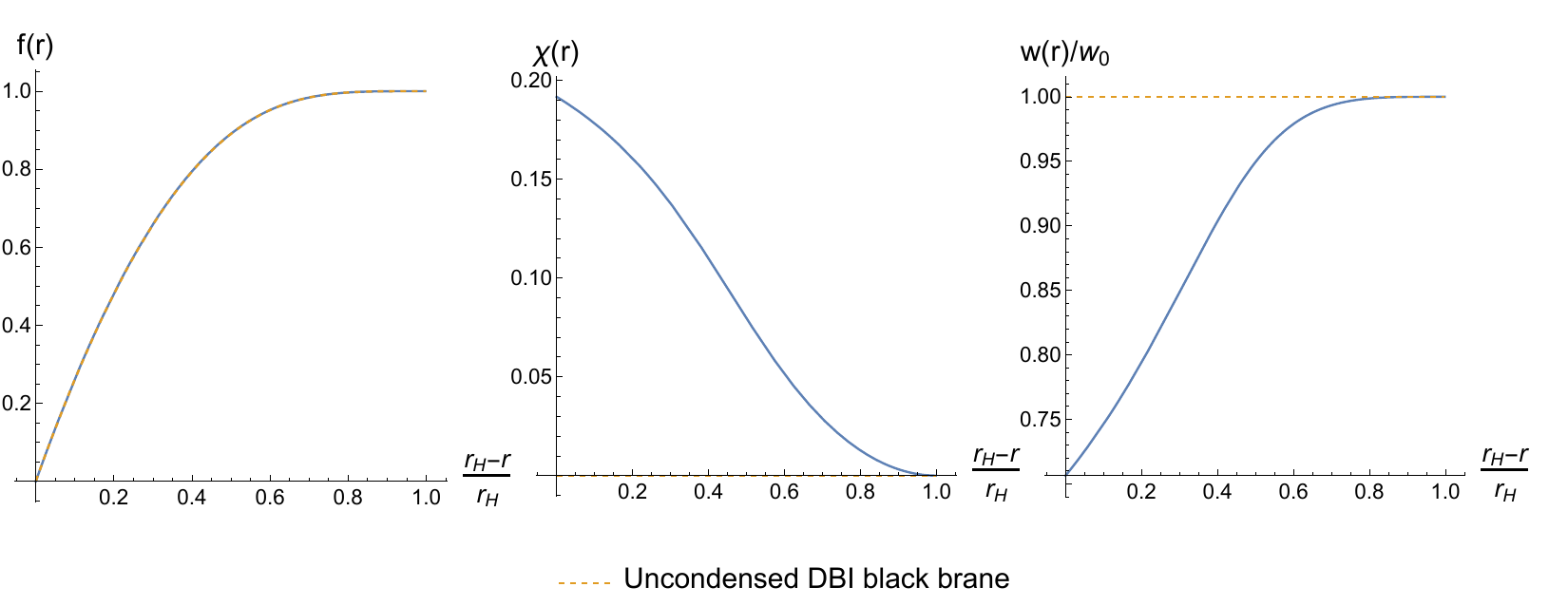}
\caption{Profiles of the bulk fields in the Higgsed solution, for $m_{WZ}^2=4m_{UV}^2$,$w_0=x_f=1$ and $\mathcal{T} = \mathcal{T}_c/2$ ($\mathcal{T}\equiv T/\hat{n}^{1/3}$). We use the coordinate $v\equiv(r_H-r)/r_H$, which goes from 0 at the horizon to 1 at the boundary. Shown are the blackening function (left), the condensed scalar (middle) and the color flux (right). The orange dashed line indicates the corresponding profiles in the uncondensed DBI black brane, which is nearly confounded with the condensed solution for $f(r)$.}
\label{fh1}
\end{center}
\end{figure} 

With the potentials \eqref{h1}, the model contains three parameters: $m_{WZ}^2$, $x_f$ and $w_0$. For given parameters, the equations of motion \eqref{E4}-\eqref{E8} may be solved numerically, with the regular horizon boundary conditions presented in section \ref{S121}. As a result, we obtain the bulk profiles of the ansatz fields, which depend on the ratio of temperature to density, $\mathcal{T}\equiv T/\hat{n}^{1/3}$. An example is shown in figure \ref{fh1}, where we plot the bulk fields for $m_{WZ}^2=4m_{UV}^2$, $w_0=x_f=1$ and $\mathcal{T} = \mathcal{T}_c/2$ (with $\mathcal{T}_c$ the transition temperature). Note that $f(r)$ (left figure \ref{fh1}) is almost identical to the uncondensed solution (the relative difference is less than 0.4\%), which is also true for the scale factor $A(r)$. This is a consequence of the small values taken by the scalar hair (middle figure \ref{fh1}).

\begin{figure}[h!]
\begin{center}
\includegraphics[scale=0.83]{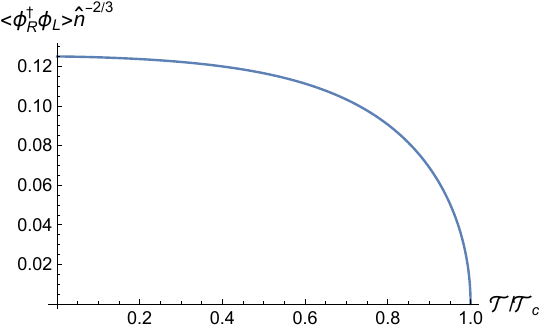}\hspace{-0.3cm}
\includegraphics[scale=0.83]{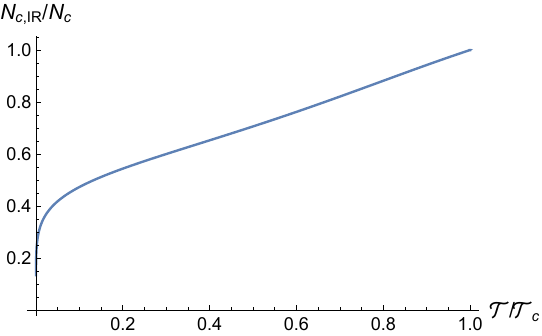}
\caption{The di-squark condensate and IR effective number of colors as a function of temperature in units of density $\mathcal{T}\equiv T/\hat{n}^{1/3}$, for $m_{WZ}^2=4m_{UV}^2$ and $w_0=x_f=1$, with $\mathcal{T}_c$ the critical temperature. The condensate is also measured in units of the density $\hat{n}$.}
\label{fh2}
\end{center}
\end{figure} 

The existence of a non-trivial scalar hair $\chi(r)$, is in direct correspondence with the paired nature of the solution. The latter implies a non-zero chiral condensate at the boundary, which depends on temperature as shown in left figure \ref{fh2}. The qualitative behavior of the condensate remains essentially the same for different parameters. In particular, it obeys the mean-field scaling near the transition
\begin{equation}
\label{h3} \left<\phi_R^\dagger\phi_L\right> \underset{\mathcal{T}\to \mathcal{T}_c}{\sim} (\mathcal{T}_c-\mathcal{T})^{1/2} \, .
\end{equation} 

What distinguishes the solution from usual chiral symmetry breaking, is the Higgsing of the gauge group, which is visible from the decrease of $w(r)$ in the IR (right figure \ref{fh1}). {From the string theory point of view, the profile of $w(r)$ describes the distribution of color branes nucleated in the bulk. In several top-down examples \cite{Yamada:2008em,McInnes:2009zp,Herzog:2009gd,Henriksson:2019zph},}{brane nucleation was seen to be accompanied by a runaway behavior of the Seiberg-Witten type \cite{Seiberg:1999xz},}{ with the branes moving off to the boundary. As clearly visible on right figure \ref{fh1},} {the Higgsed color branes in our model instead spread over a finite distance from the horizon, controlled by the chemical potential. Moreover, this solution is expected to be stable since it is a priori the unique condensed solution\footnote{{Except for solutions with a higher number of nodes in the scalar field profile, which are generally expected to be unstable. Inhomogeneous solutions could also compete in principle, but are not expected to arise in absence of a Chern-Simons term.}}. There is therefore no indication of the runaway behavior observed in top-down examples. Note that the same was observed in the D3-D7 setup of \cite{Faedo:2018fjw}}{ (summarized in appendix \ref{AppA1}),}{where the location of the instanton describing the dissolved color brane was controlled by the isospin chemical potential.}

The effective number of colors in the IR is given by
\begin{equation}
\label{h4} N_{c,IR} = N_c w(r_H) \, ,
\end{equation}  
which decreases with temperature as shown in figure \ref{fh2} (right). This behavior is qualitatively different from an $N_c = 3$ CFL phase, for which the gauge group would be fully Higgsed below the transition temperature. The progressive Higgsing observed in figure \ref{fh2} might be a general property of large $N_c$ color superconductors {featuring a continuous phase transition}, which would be interesting to investigate further.  

Another important observation from figure \ref{fh2} is that, even at zero temperature, the gauge group is not fully Higgsed. More precisely, $N_{c,IR}(T=0)$ is about 10\% of $N_c$, meaning that most of the color group is Higgsed in the IR, but there is still some residual gauge subgroup. Interestingly, the analysis below shows that this conclusion remains unchanged for different model parameters.

\subsubsection{Parameter dependence}

We will now describe the qualitative change in the condensed solution upon variation of the model parameters. This analysis is divided into separate discussions for each of the parameters $x_f$, $m_{WZ}^2$ and $w_0$, and focused on zero temperature solutions. The case of finite temperature is qualitatively similar, but we found the parameter dependence to be clearest at $T=0$. 

\begin{figure}[h!]
\begin{center}
\includegraphics[scale=0.83]{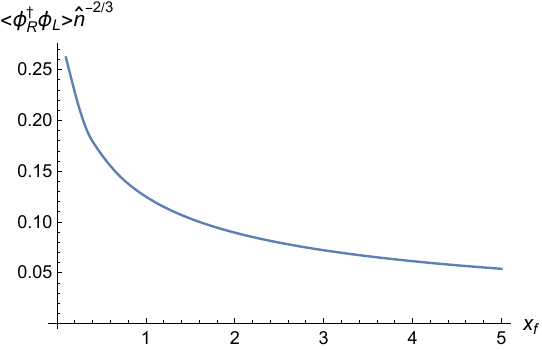}\hspace{-0.3cm}
\includegraphics[scale=0.83]{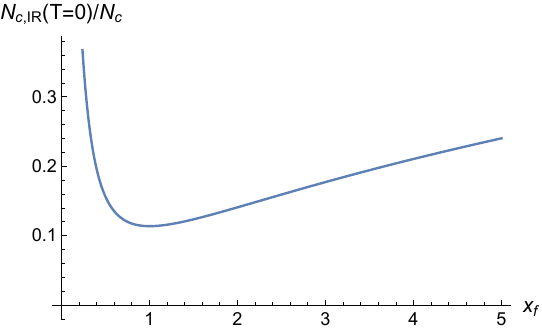}
\caption{The di-squark condensate and IR effective number of colors as a function of $x_f$, at $m_{WZ}^2=4m_{UV}^2$,$w_0=1$ and $T=0$.}
\label{fh3}
\end{center}
\end{figure}

We start by discussing the dependence on the parameter $x_f$, which is proportional to the ratio of flavor to color number $N_f/N_c$. Figure \ref{fh3} shows the zero-temperature condensate and IR color number $N_{c,IR}$, as a function of $x_f$. At low $x_f$, the Higgsing is observed to become stronger with increasing $x_f$, which is a behavior similar to color-flavor locking.  However, $N_{c,IR}$ reaches a minimum at $x_f\simeq 1$, beyond which it increases with $x_f$. Unlike the Higgsing fraction, the condensate in left figure \ref{fh3} shows a monotonous decrease, without distinction between small and large $x_f$. 

\begin{figure}[h!]
\begin{center}
\includegraphics[scale=0.82]{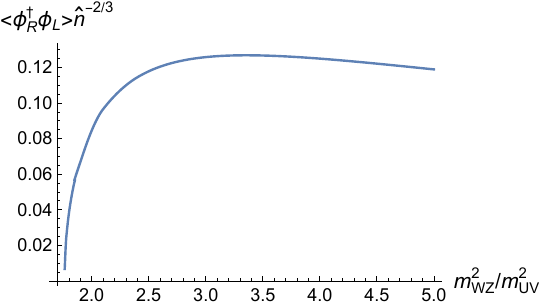}\hspace{-0.5cm}
\includegraphics[scale=0.86]{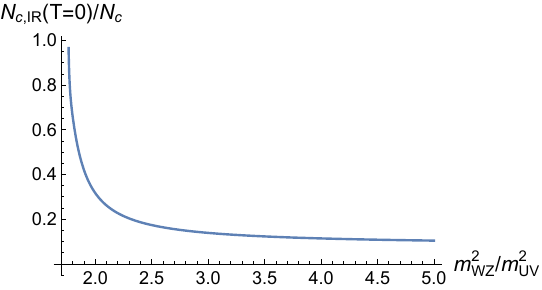}
\caption{The di-squark condensate and IR effective number of colors as a function of $m_{WZ}^2/m_{UV}^2$, at $x_f=w_0=1$ and $T=0$. The critical value where the phase transition occurs is at $m_{WZ}^2/m_{UV}^2\simeq 1.76$.}
\label{fh4}
\end{center}
\end{figure}

The dependence on the Wess-Zumino mass $m_{WZ}^2$ can be seen from figure \ref{fh4}. The latter shows the change in the di-squark condensate and the IR color number $N_{c,IR}(T=0)$, for different values of $m_{WZ}^2$. Equation \eqref{E10} would naively suggest that increasing $m_{WZ}^2$ should make the Higgsing stronger. However, figure \ref{fh4} (right) shows that the size of the IR gauge group is nearly independent of $m_{WZ}^2$, except very close to the phase transition. The condensate in left figure \ref{fh4} follows the same type of behavior, with the dependence on $m_{WZ}^2$ concentrated near the transition. 


\begin{figure}[h!]
\begin{center}
\includegraphics[scale=0.82]{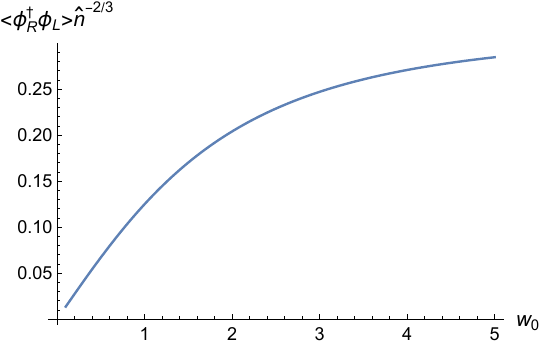}\hspace{-0.3cm}
\includegraphics[scale=0.84]{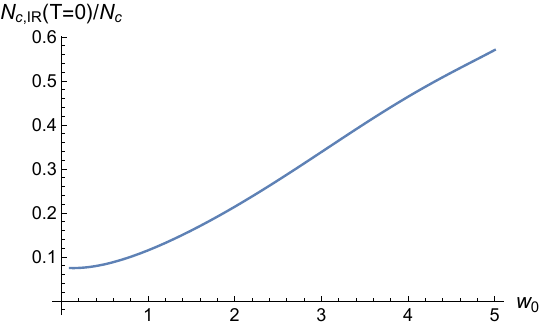}
\caption{The di-squark condensate and IR effective number of colors as a function of $w_0$, at $m_{WZ}^2=4m_{UV}^2$,$x_f=1$ and $T=0$.}
\label{fh6}
\end{center}
\end{figure}

The last parameter $w_0$ controls the color brane tension. Equation \eqref{E10} would again suggest that a larger fraction of the gauge group should be Higgsed at low $w_0$. This is indeed what is observed on figure \ref{fh6}, but the zero temperature IR color number goes to a finite value as $w_0$ approaches 0. Overall, both the condensate and the rank of the IR gauge group increase with $w_0$.  

An important observation from figures \ref{fh3},\ref{fh4} and \ref{fh6}, is that $N_{c,IR}$ remains self-consistently positive for any values of the parameters. This is an interesting property of our solution, which however also implies that realizing full Higgsing is hard in this model, {although it could be approached with a different choice of potentials. In particular, the analysis of the next subsection suggests that potentials associated with a first order phase transition generate stronger Higgsing. }

\begin{figure}[h!]
\begin{center}
\includegraphics[scale=0.9]{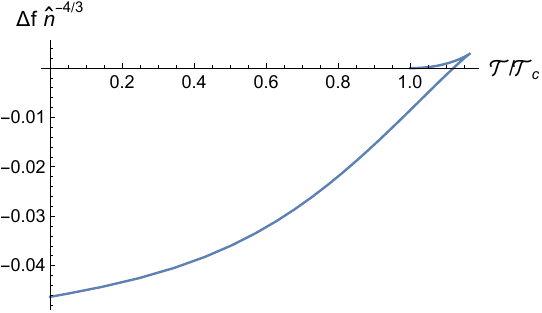}
\caption{{Free energy difference of the hairy solution with the uncondensed DBI black brane, in units of density $\hat{n}$ and as a function of temperature $\mathcal{T}\equiv T/\hat{n}^{1/3}$, for the potentials \eqref{h5}.} {The parameters are set to $w_0=x_f=1$, $m_{WZ}^2 = 4m_{UV}^2$ and $\a = 8$. $\mathcal{T}_c$ is the temperature where the DBI black brane becomes perturbatively unstable (which can be read from figure \ref{fpd1},}{ and is smaller than the actual transition temperature in the first order case).}}
\label{fh7}
\end{center}
\end{figure}

{
\subsection{First order transition}

\label{S32}

The simple quadratic potentials analyzed in the previous subsection were found to always generate a second order phase transition. Here, we are interested in potentials for which the transition becomes first order. Studying such potentials is interesting both for a broader understanding of the model, and for getting closer to the expected behavior of the CFL transition at $N_c=N_f=3$ \cite{Alford:2001dt}.}{We found that a first order transition can arise for example upon adding a quartic piece to the quadratic WZ potential \eqref{h1}}, {so that the flavor potentials take the form
\begin{equation}
\label{h5} \ell^2V_f(\chi) = 1+\frac{1}{2}m_f^2\ell^2 \chi^2 \sp \ell^2V_{WZ}(\chi) = \frac{1}{2w_0}m_{WZ}^2\ell^2 \chi^2\left(1+ \a \chi^2\right) \, . 
\end{equation} 
With these potentials, and fixing the model parameters to $w_0=x_f=1$ and $m_{WZ}^2 = 4m_{UV}^2$, we found that the transition becomes first order for $\alpha$ larger than some $\a_1$, estimated numerically to lie around $\a_1 \simeq 5$. For illustration, figure \ref{fh7}}{ shows the free energy difference with the DBI black brane as a function of temperature for $\a = 8$.} 

\begin{figure}[h!]
\begin{center}
\includegraphics[scale=0.55]{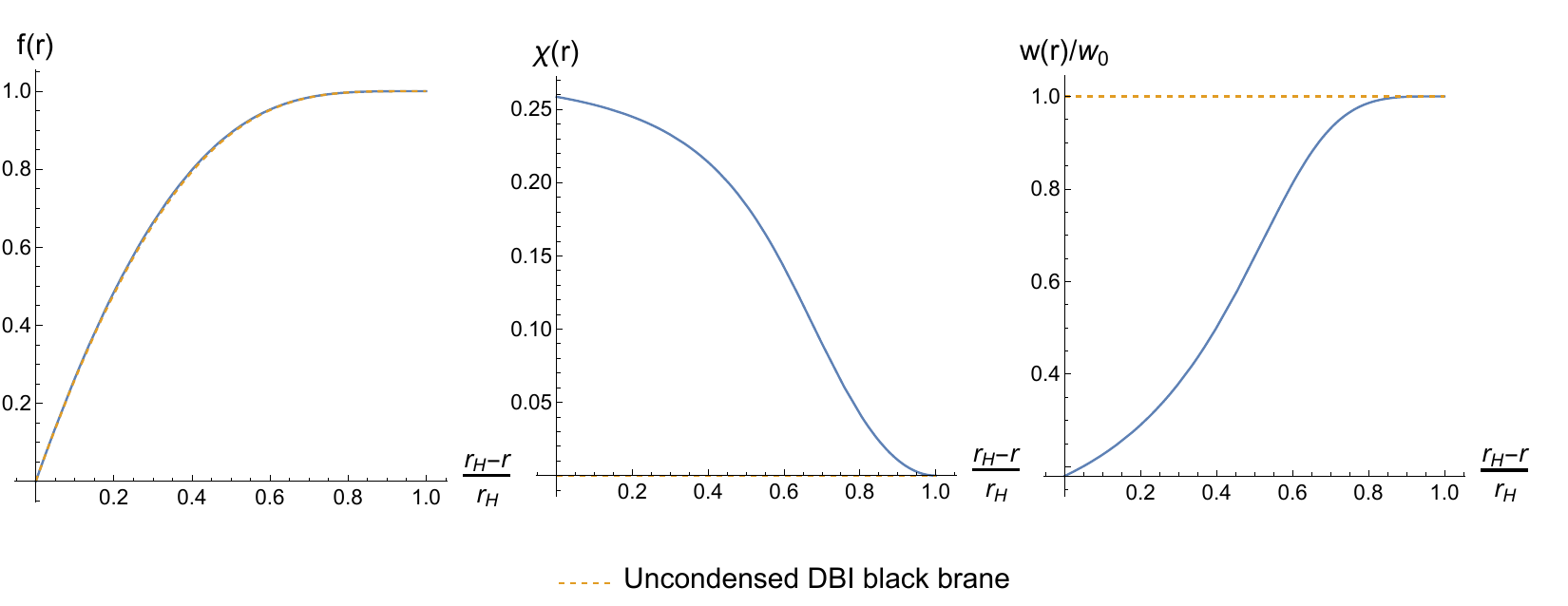}
\caption{{Same as figure \ref{fh1},}{ but for the potentials \eqref{h5}}{with parameter $m_{WZ}^2=4m_{UV}^2$, $w_0=x_f=1$ and $\a=8$, and at $\mathcal{T} = \mathcal{T}_c/2$. See the caption of figure \ref{fh7}}{ for the precise definition of the critical temperature $\mathcal{T}_c$.}}
\label{fh1b}
\end{center}
\end{figure} 

{To see how the CSC solution is modified by the change of potential, we show in figure \ref{fh1b}}{ the bulk fields for the leading hairy black brane state at $\mathcal{T}=\mathcal{T}_c/2$ ($\mathcal{T}\equiv T/\hat{n}^{1/3}$). Comparing with figure \ref{fh1}} {does not reveal any significant qualitative difference. At the quantitative level however, the scalar hair in figure \ref{fh1b}}{is seen to be significantly larger compared with figure \ref{fh1}.}{ This is also associated with a much stronger Higgsing in the first order case, as can be seen from right figure \ref{fh1b}.}{ Left figure \ref{fh1b}}{ indicates that the back-reaction of the hair on the geometry is still small, but larger than for potentials \eqref{h1}} {(a more detailed analysis shows that the relative difference of $f(r)$ and $A(r)$ with the DBI black brane reaches around $1\%$). } 
	
\begin{figure}[h!]
\begin{center}
\includegraphics[scale=0.83]{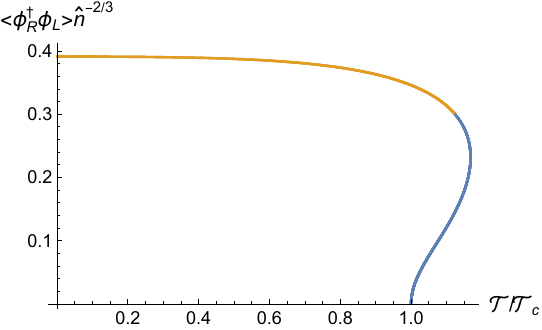}\hspace{-0.3cm}
\includegraphics[scale=0.83]{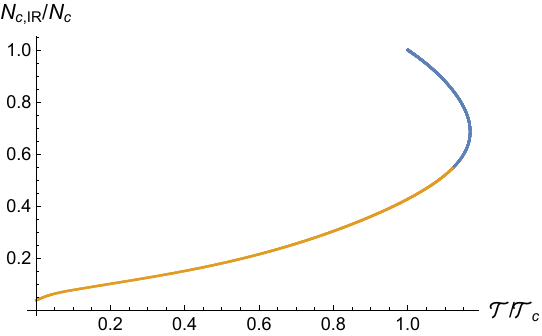}
\caption{{The di-squark condensate and IR effective number of colors as a function of temperature in units of density $\mathcal{T}\equiv T/\hat{n}^{1/3}$, for the potentials \eqref{h5},}{with parameters $m_{WZ}^2=4m_{UV}^2$, $w_0=x_f=1$ and $\a=8$. The condensate is also measured in units of the density $\hat{n}$. The orange part of the line indicates where the condensed solution dominates over the DBI black brane.}}
\label{fh2b}
\end{center}
\end{figure}

{For completeness, figure \ref{fh2b}}{ shows the scalar condensate and the size of the IR color group as a function of temperature. As expected from the analysis of the solution, comparison with figure \ref{fh2}}{ shows that both the condensate and the Higgsing fraction are significantly larger in the first order case. In particular, the zero temperature solution is very close to full Higgsing, with more than $96\%$ of the gauge group Higgsed. 
	
The previous quantitative statements are of course parameter-dependent, but this single example already indicates the trend induced by the first order transition.

As a final note, let us mention another interesting feature of the first order solution, related to the chiral symmetry breaking fraction $x_b$. As discussed in appendix \ref{AppD},}{ the fully broken solution with $x_b=1$ always dominates for the quadratic potentials \eqref{h1}.}{ This is no longer the case however with the potentials \eqref{h5},}{ for which the leading solution has $x_b = x_b^*$, where the fraction $x_b^*$ depends on model parameters but not on temperature (and solutions with $x_b<x_b^*$ are unstable). In particular, for the parameters chosen above, $w_0=x_f=1$, $m_{WZ}^2 = 4m_{UV}^2$ and $\a=8$, a numerical calculation gives $x_b^* \simeq 0.875$. Note that this means that the solution shown above in figures \ref{fh7}, \ref{fh1b} and \ref{fh2b},}{ which has $x_b=1$, is not the true ground state of the system. Its properties are however very similar to the solution with $x_b = x_b^*$ (at least for this choice of parameters), so the simpler case $x_b=1$ gives a good idea of how the first order solution behaves. 
	
}

\section*{Acknowledgements}\label{ACKNOWL}
\addcontentsline{toc}{section}{Acknowledgements}

I would like to thank Javier Subils and Tuna Demircik for discussion. I am also grateful to Elias Kiritsis, Francesco Nitti and Matti J\"arvinen for comments on a draft. 

This work is supported by the Netherlands Organisation for Scientific Research (NWO) under the VICI grant VI.C.202.104.

\newpage

\appendix
\renewcommand{\theequation}{\thesection.\arabic{equation}}
\addcontentsline{toc}{section}{Appendix\label{app}}
\section*{Appendix}

\section{Lessons from a top-down example}
\label{AppA1}

In this appendix, we analyze the reduced theory that arises from projecting the setup of \cite{Faedo:2018fjw} to five dimensions. The result gives useful indications for constructing a bottom-up color superconductor.

The model considered in \cite{Faedo:2018fjw} is based on the type IIB D3/D7 configuration \cite{Karch:2002sh}, which is dual to $\mathcal{N}=4$ super-Yang-Mills coupled to $N_f$ fundamental hypermultiplets. In this setup, a stack of $N_f$ probe D7 flavor branes are placed in the $AdS_5\times S^5$ background generated by a stack of $N_c$ D3 color branes. The bulk ten-dimensional metric is expressed by
\begin{equation}
\label{A11} \intd s^2 = H^{-\frac{1}{2}}\left(-\intd t^2 + \intd \vec{x}^2\right) + H^{\frac{1}{2}} \left(\intd y_i^2 + \intd z_\a^2\right)\, ,
\end{equation}
with $(t,\vec{x})$ the boundary coordinates, $(y_i)_{1\leq i\leq 4}$ the coordinates along the D7 branes that are transverse to the D3 branes, and $(z_\a)_{\a=1,2}$ the coordinates transverse to both sets of branes. In spherical coordinates, the metric in the $y_i$ directions takes the form
\begin{equation}
\label{A12} \intd y_i^2 = \intd r^2 + r^2(\omega_1^2+\omega_2^2+\omega_3^2) \, ,
\end{equation} 
where the $\omega_i$'s are the left-invariant forms on $S^3$. $H$ is the harmonic function in the six-dimensional space spanned by $y_i$ and $z_\a$
\begin{equation}
\label{A13} H=\frac{\ell^4}{(r^2+z_1^2+z_2^2)^2}\, ,
\end{equation}
with $\ell$ the AdS radius. The background also contains a RR 4-form $C_4$, whose expression is given by
\begin{equation}
\label{A14} C_4 = - H^{-1} \intd t \wedge \intd x^3
\end{equation}

The bulk action takes a similar form to \eqref{A1}
\begin{equation}
\label{A15} S = S_c + S_f + S_{WZ} \,,
\end{equation}
where the color part $S_c$ (which corresponds to type IIB supergravity) determines the background solution above, whereas $S_f$ and $S_{WZ}$ fully control the flavor dynamics in the probe limit $N_f\ll N_c$. 

The flavor part of the action controls the dynamics of the D7-branes massless worldvolume fields, including a $U(N_f)$ gauge field $A_\mu^a$, and two adjoint scalars $Z^\a_a$, where $a$ is the adjoint flavor index. $S_f$ is the low-energy action for the D7 branes, which is given by the so-called super-Yang-Mills-Higgs (SYMH) action 
\begin{equation}
\label{A16} S_f = -\frac{1}{2}\int\mathrm{Tr}\left(F\wedge\star F + H^{\frac{1}{2}}\d_{\a\b}DZ^\a\wedge\star DZ^\b\right) \, .
\end{equation}
The flavor action also contains a topological WZ part $S_{WZ}$, which couples the flavor fields to the background RR 4-form
\begin{equation}
\label{A17} S_{WZ} = \frac{1}{2}\int C_4\wedge \mathrm{Tr}(F\wedge F) \, .
\end{equation}

Equation \eqref{A17} makes it clear that a worldvolume instanton induces color charge on the flavor branes. In other words, a bulk solution with non-zero instanton number for the flavor gauge fields corresponds to a Higgsed state. Accordingly, the color superconductor of \cite{Faedo:2018fjw} is given by such an instanton solution, whose size is dynamically generated at finite isospin density. The solution takes the form\footnote{For simplicity, \cite{Faedo:2018fjw} fixes the number of flavors to $N_f=2$.}
\begin{equation}
\label{A18} A \equiv A_\mu^a \intd x^\mu \otimes \s_a = a_t(r)\intd t\otimes \s_3 + a(r)\d^{ab}\omega_a\otimes \s_b \, ,
\end{equation}
\begin{equation}
\label{A19} Z^1 = \phi(r)\s_3 \sp Z^2 = 0 \, ,
\end{equation}
with $\s_a$ the Pauli matrices. 


We are interested in the five-dimensional action that describes the dynamics of the ansatz fields \eqref{A18}-\eqref{A19}. The latter can be computed by substituting the ansatz into \eqref{A16} and \eqref{A17}, and performing the integral over the $S^3$ directions. As a result, we get
\begin{align}
\nn S_{f,5} = 12\pi^2\int\intd t\intd x^3\intd r r^3\bigg[&\frac{1}{3}a_t'(r)^2-\frac{a'(r)^2}{Hr^2}+\frac{8a(r)^2a_t(r)^2}{3r^2}-\frac{4a(r)^2(1+a(r))^2}{Hr^4}\\
\label{A110} &-\frac{1}{6}\phi'(r)^2 - \frac{8a(r)^2\phi(r)^2}{r^2} \bigg] \, , 
\end{align}
\begin{equation}
\label{A111} S_{WZ,5} = 24\pi^2 \int C_4 \wedge \intd V(a) \, ,
\end{equation}
with 
\begin{equation}
\label{A112} V(a) = a^2 \left(1+\frac{2}{3}a\right) \, .
\end{equation}
This setup therefore provides a concrete example where a WZ term of the form that is considered in this work \eqref{A5} arises.  

The kinetic part of the action \eqref{A110} involves three fields: the scalar $a$, which is dual to the di-squark operator \cite{Faedo:2018fjw}, the isospin gauge field $a_t$, and an additional scalar $\phi$ which descends from the Higgs field $Z^\a$. Apart from the presence of this additional field $\phi$, the main difference with our model (at quadratic order) is the direct coupling between $a$ and $a_t$, which generates a mass for $a$ proportional to $-a_t^2$. This term arises because the instanton is charged under isospin. 

In this setup, the condensation of the scalar at finite isospin density therefore occurs via the standard holographic superconductor mechanism \cite{Gubser08a}. It is thus crucial for isospin density to be turned on, instead of baryon density. In order to investigate the generation of color superconducting phases at finite baryon density, it is necessary to consider higher derivative corrections to the flavor action \eqref{A16}.

\section{Equations of motion}
\label{AppA}

We present in this appendix the general equations of motion derived from the bulk action \eqref{A1}. These equations are conveniently written in terms of the effective metric on the flavor branes
\begin{equation}
	\label{eom0} \bar{g}_{MN} \equiv g_{MN} + \ell^2F_{MN} + \ell^2\pa_M\chi\pa_N\chi \, ,
\end{equation}
with the convention that raised indices refer to the inverse of $\bar{g}$, rather than contraction with the inverse of the metric $g$
\begin{equation}
	\label{eom0b} \bar{g}^{MN} \equiv (\bar{g}^{-1})_{MN} \neq g^{MP}g^{NQ}\bar{g}_{PQ} \, .
\end{equation}

With these definitions, the equation for the $U(1)$ gauge field is
\begin{equation}
	\label{eom1} \pa_M\left(\sqrt{-\bar{g}}V_f(\chi)\bar{g}^{[MN]}\right) = 0 \, ,
\end{equation}
for the RR 4-form 
\begin{equation}
	\label{eom2} \nabla_M F_5^{MNPQR} = -\frac{x_f}{\sqrt{-g}}\ell V_{WZ}'(\chi)  \, \e^{MNPQR}\pa_M\chi \, ,
\end{equation}
and for the scalar
\begin{equation}
	\label{eom3} \frac{1}{\sqrt{-\bar{g}}V_f(\chi)}\pa_M\left(\sqrt{-\bar{g}}V_f(\chi)\bar{g}^{(MN)}\pa_N\chi\right) - \frac{V_f'(\chi)}{V_f(\chi)}+ \frac{V_{WZ}'(\chi)}{V_f(\chi)}\sqrt{\frac{g}{\bar{g}}}\ell(\star F_5) = 0 \, .
\end{equation}
We used the standard notations $[MN]$ and $(MN)$ for the antisymmetric and symmetric parts.

The Einstein equations take the usual form
\begin{equation}
	\label{eom4} R_{MN} - \frac{1}{2}(R-\L)g_{MN} = \frac{1}{2}\left(T^f_{MN}+T^5_{MN}\right) \, ,
\end{equation}
with the stress-energy tensors given by
\begin{equation}
	\label{eom5} T^f_{MN} = -\frac{N_f}{N_c}\sqrt{\frac{\bar{g}}{g}}V_f(\chi)\bar{g}^{(PQ)}g_{MP}g_{NQ} \sp T^5_{MN} = \frac{1}{4!} F_{MPQRS}F_N^{\hp{N}PQRS} - \frac{1}{2.5!}F_5^2 g_{MN} \, .
\end{equation}

\section{Details on the phase diagram}
\label{AppB}

In this appendix, we provide detailed derivations of some of the limiting properties of the phase diagram, which is presented in section \ref{S22}. Specifically, we consider the two limits $m_{WZ}^2/m_{UV}^2\gg 1$ and $T\to 0$.

\subsection{The large $m_{WZ}^2$ limit}

Figure \ref{fpd1} indicates that the critical temperature $\mathcal{T}_c$ ($\mathcal{T}\equiv T/\hat{n}^{1/3}$) keeps increasing as $m_{WZ}^2$ gets more negative. When $\mathcal{T}_c$ becomes much larger than 1, the horizon value $u_{H,c}$ of the coordinate $u$ (defined in \eqref{ms9}) at the transition can be expressed from \eqref{u12}   
\begin{equation}
\label{B1} u_{H,c} = \frac{1}{\pi\mathcal{T}_c}\left[ 1+\OO\left(\mathcal{T}_c^{-6}\right) \right] \, .
\end{equation}
Since $u_{H,c}$ goes to zero, when the transition happens, the coordinate $u$ is much smaller than 1 over the entire bulk. In this limit, the effective mass squared \eqref{ms3} simplifies to 
\begin{equation}
\label{B2} m_{eff}^2(u) = m_{UV}^2 \left[1+\frac{m_{WZ}^2}{2m_{UV}^2}u^6\left(1+\OO\left(\frac{m_{UV}^2}{m_{WZ}^2}\right)+\OO(u^6)\right)\right] \, .
\end{equation}
$m_{eff}^2$ then becomes of a function of the variable
\begin{equation}
\label{B3} v\equiv \frac{m_{WZ}^2}{m_{UV}^2}u^6 \, ,
\end{equation}
and $m_{eff}^2(v)$ does not have any parameter. As a result, the phase transition occurs for a given value of $v$ at the horizon, denoted $v_{H,c}$. Now substituting the expression of $v$ \eqref{B3}, we finally arrive at a relation that determines the behavior of the critical temperature
\begin{equation}
\label{B4} v_{H,c} = \frac{m_{WZ}^2}{m_{UV}^2}u_{H,c}^6 = \frac{m_{WZ}^2}{m_{UV}^2} (\pi\mathcal{T}_c)^{-6}\left[ 1+\OO\left(\mathcal{T}_c^{-6}\right) \right]\, ,
\end{equation}  
where we also used \eqref{B1}. Inverting \eqref{B4} gives
\begin{equation}
\label{B5} \mathcal{T}_c = \pi^{-1}\left(v_{H,c}^{-1}\frac{m_{WZ}^2}{m_{UV}^2}\right)^{1/6}\left[ 1+\OO\left(\frac{m_{UV}^2}{m_{WZ}^2}\right) \right]\, .
\end{equation}

\subsection{The extremal $AdS_2$ geometry}

We will now provide some details about the $AdS_2$ geometry that arises near the extremal horizon of the DBI black brane solution (presented in section \ref{S13}) at zero temperature. In particular, we will derive the relation between the $AdS_2$ length $\ell_2$ and the $AdS_5$ length $\ell$.

Recall that the metric of the solution takes the form 
\begin{equation}
\label{B6} \intd s^2 = \ex^{2A(r)} \left(-f(r)\intd t^2 + f(r)^{-1}\intd r^2 + \intd\vec{x}^2\right) \, , 
\end{equation} 
where $A(r)=-\log(r/\ell)$ and the blackening function $f(r)$ vanishes at the horizon radius $r_H$. In the extremal limit $(r_H\to r_e)$, the first derivative $f'(r_e)$ also goes to zero, such that the expansion of $f(r)$ near $r_e$ takes the form
\begin{equation}
\label{B7} f(r) = \frac{1}{2} f''(r_e) \left(r-r_e\right)^2 + \OO(r-r_e)^3 \, .
\end{equation} 

The near-horizon limit of the geometry \eqref{B6} can then be obtained by considering
\begin{equation}
\label{B8} r = r_e \left(1- \e \frac{r_e\ell_2}{\ell^2}\frac{\ell_2}{\zeta}\right) \sp t = \e^{-1} \tau \sp \e \ll 1 \, ,
\end{equation}
where $\zeta$ is the $AdS_2$ radial coordinate, $\tau$ the time coordinate, and the expression of $\ell_2$ is to be determined. In the limit \eqref{B8}, the metric \eqref{B6} takes the form
\begin{equation}
\label{B9} \intd s^2 = \left[\intd s_2^2 + \left(\frac{\ell}{r_e}\right)^2\intd\vec{x}^2\right]\left[1+\OO(\e)\right] \, , 
\end{equation}
\begin{equation}
\label{B10} \intd s_2^2 = \left(\frac{\ell_2}{\zeta}\right)^2 \left(-\frac{1}{2}r_e^2f''(r_e)\left(\frac{\ell_2}{\ell}\right)^2\intd \tau^2 + \frac{2}{r_e^2f''(r_e)}\left(\frac{\ell_2}{\ell}\right)^{-2}\intd \zeta^2\right) \, .
\end{equation}
Equation \eqref{B10} is nothing but the $AdS_2$ metric in Poincar\'e coordinates, if we make the identification
\begin{equation}
\label{B11} \frac{\ell_2}{\ell} = \left(\frac{1}{2}r_e^2f''(r_e)\right)^{-\frac{1}{2}} \, .
\end{equation}
This relation, between the $AdS_2$ length and the second derivative of the blackening function at the horizon, generally applies to any extremal black brane geometry with $A(r)=-\log(r/\ell)$. For the DBI black brane, the blackening function is given by \eqref{u9}, from which $\ell_2$ can be computed to be 
\begin{equation}
\label{B12} \frac{\ell_2}{\ell} = \frac{1}{2\sqrt{3}}\sqrt{\frac{12+x_f}{6+x_f}} \, .
\end{equation}

\section{Partial breaking of the chiral symmetry}

\label{AppD}

In this appendix, we consider a more general ansatz for the bulk tachyon field dual to the di-squark operator, allowing for chiral symmetry to be partially broken 
\begin{equation}
\label{D1} T = \chi \mathbb{I}_{N_b} \sp N_b \leq N_f \, .
\end{equation}
For this ansatz, the bulk flavor action takes the form 
\begin{align}
\nn S_f =  &-N_c^2x_fx_b\ell^{-3} \int\intd^5x\, V_f(\chi)\sqrt{-\det{(g_{MN}+\ell^2F_{MN}+\ell^2\pa_M\chi\pa_N\chi)}} \\
\label{D2} &-N_c^2x_f(1-x_b)\ell^{-3} \int\intd^5x\, V_f(0)\sqrt{-\det{(g_{MN}+\ell^2F_{MN})}} \, ,
\end{align}
where we defined 
\begin{equation}
\label{D3} x_b \equiv \frac{N_b}{N_f} \, ,
\end{equation}
whereas the WZ action is only modified by a factor $x_b$
\begin{equation}
\label{D4} S_{WZ} = N_c^2x_fx_b\ell^{-2} \int C_4\wedge \intd V_{WZ}(\chi) \, .
\end{equation}
Note that, whereas $x_f$ is a model parameter, $x_b$ is rather a parameter of the solution. In particular, at equilibrium, $x_b$ will take the value that minimizes the free energy. 

The equations of motion for a solution with a general symmetry breaking fraction are derived from \eqref{D1} and \eqref{D2} (together with the color action \eqref{A2}). The scalar equation is still given by \eqref{eom3} 
\begin{equation}
\label{D4b} \frac{1}{\sqrt{-\bar{g}}V_f(\chi)}\pa_M\left(\sqrt{-\bar{g}}V_f(\chi)\bar{g}^{(MN)}\pa_N\chi\right) - \frac{V_f'(\chi)}{V_f(\chi)}+ \frac{V_{WZ}'(\chi)}{V_f(\chi)}\sqrt{\frac{g}{\bar{g}}}\ell(\star F_5) = 0 \, ,
\end{equation}
but all the other equations are modified. The equation for the 4-form only receives a factor $x_b$
\begin{equation}
\label{D5} \nabla_MF_5^{MNPQR} = -\frac{x_fx_b}{\sqrt{-g}}\ell V_{WZ}'(\chi) \e^{MNPQR}\pa_M\chi \, ,
\end{equation}
whereas the $U(1)$ gauge field obeys 
\begin{equation}
\label{D6} \pa_M\left(x_b \sqrt{-\bar{g}}V_f(\chi)\bar{g}^{[MN]} + (1-x_b)\sqrt{-\bar{g}_{(0)}}V_f(0)\bar{g}_{(0)}^{[MN]}\right) \, ,
\end{equation}
with the effective metrics defined as
\begin{equation}
\label{D7} \bar{g}_{MN}^{(0)} \equiv g_{MN} + F_{MN} \, ,
\end{equation}
\begin{equation}
\label{D8} \bar{g}_{MN} \equiv \bar{g}_{MN}^{(0)} + \pa_M\chi \pa_N\chi \, .
\end{equation}
In the Einstein equations, the flavor stress-tensor is now the sum of a broken and an unbroken contributions
\begin{equation}
\label{D9} T^f_{MN} = -x_fg_{MP}g_{NQ}\left(x_b\sqrt{\frac{\bar{g}}{g}}V_f(\chi)\bar{g}^{(PQ)}+ (1-x_b)\sqrt{\frac{\bar{g}_{(0)}}{g}}V_f(0)\bar{g}_{(0)}^{(PQ)} \right) \, .
\end{equation}

We now substitute the black brane ansatz \eqref{E1}-\eqref{E3} into the equations above. For the 4-form, we get
\begin{equation}
\label{D10} w' = x_fx_b \ell^2 \pa_r V_{WZ}(\chi) \, ,
\end{equation}
whereas the gauge field equation of motion is given by
\begin{equation}
\label{D11} \pa_r\left[\ex^A A_t'\left(x_b\Xi^{-\frac{1}{2}}V_f(\chi) + (1-x_b)G^{-\frac{1}{2}}V_f(0)\right)\right] \, ,
\end{equation}
where we recall the definitions
\begin{equation}
\label{D12} G(r) \equiv 1 - \ex^{-4A(r)}\ell^4 A_t'(r)^2 \, ,
\end{equation}
\begin{equation}
\label{D13} \Xi(r) \equiv G(r) + \ex^{-2A(r)}f(r)\ell^2\chi'(r)^2 \, .
\end{equation}
Finally, the scalar and metric equations are
\begin{equation}
\label{D13b} \ex^{-5A}\pa_r\left(\ex^{3A}f\chi'\right) + \pa_r\log{\left(\Xi^{-\frac{1}{2}}V_f(\chi)\right)}\ex^{-2A}f \chi' - \Xi \left(\frac{V_f'(\chi)}{V_f(\chi)}+ \Xi^{-\frac{1}{2}}\frac{V_{WZ}'(\chi)}{V_f(\chi)}w\right) = 0 \, ,
\end{equation}
\begin{equation}
\label{D14} A'' - (A')^2 + \frac{x_f}{6}x_b\Xi^{-\frac{1}{2}}V_f(\chi) (\chi')^2 = 0\, ,
\end{equation}
\begin{equation}
\label{D15} A'f' + 4f(A')^2 + \frac{1}{6}\ex^{2A}\left(2\Lambda+2x_f(1-x_b)V_f(0)G^{-\frac{1}{2}}+2x_fx_bV_f(\chi)\Xi^{-\frac{1}{2}}+\frac{w^2}{\ell^2}\right) \, .
\end{equation}

\begin{figure}[h!]
	\begin{center}
		\includegraphics[scale=1.1]{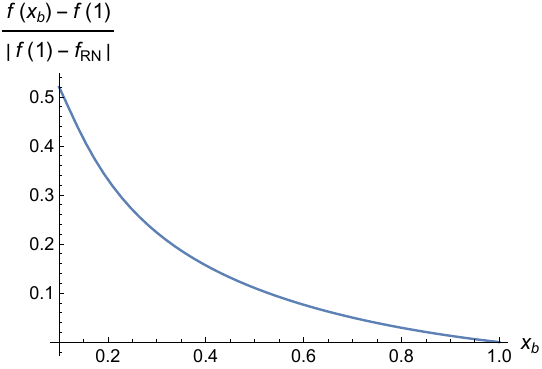}
		\caption{Free energy difference between the solution with symmetry breaking fraction $x_b$ and the fully broken solution ($x_b=1$), normalized by the difference with the RN phase at $x_b=1$. The flavor potentials were chosen as in \eqref{h1}, with parameters set to $w_0=x_f=1$ and $m_{WZ}^2=4m_{UV}^2$, whereas the temperature in units of density $\mathcal{T}\equiv T/\hat{n}^{1/3}$ is half the critical value $\mathcal{T}_c$.} 
		\label{fD1}
	\end{center}
\end{figure}

\begin{figure}[h!]
	\begin{center}
		\includegraphics[scale=1.1]{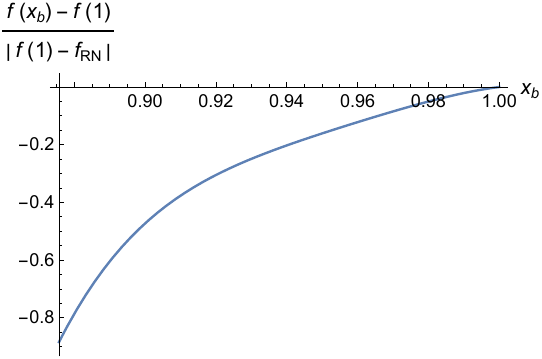}
		\caption{Same as figure \ref{fD1} but for the potentials \eqref{h5}, with parameters set to $w_0=x_f=1$, $m_{WZ}^2=4m_{UV}^2$ and $\a=8$.} 
		\label{fD2}
	\end{center}
\end{figure}

Given equations \eqref{D10}-\eqref{D15}, and the same boundary conditions presented in section \ref{S121}, the condensed solution can be obtained for any chiral symmetry breaking fraction $x_b$. Among these, the ground state corresponds to the solution with the lowest free energy. {Numerical results indicate different behaviors for the two types of potentials analyzed in this work }
\begin{itemize}
\item {For the potentials associated with a second order phase transition \eqref{h1},}{ the fully broken solution $x_b = 1$ always dominates, as can be seen in figure \ref{fD1}}{ for the case $w_0 = x_f = 1$, $m_{WZ}^2=4m_{UV}^2$ and $\mathcal{T} = \mathcal{T}_c/2$;
\item For the potentials associated with a first order transition \eqref{h5}}{ (with parameters set to $w_0 = x_f = 1$, $m_{WZ}^2=4m_{UV}^2$ and $\a=8$), the CSC solution exists only for chiral symmetry breaking fractions $x_b$ larger than some value $x_b^*\simeq 0.875$\footnote{More precisely, the CSC solution moves to high temperature $\mathcal{T}>\mathcal{T}_c$ and becomes unstable for $x_b < x_b^*$.}. Moreover, the solution with $x_b = x_b^*$ dominates for all temperatures. This is illustrated in figure \ref{fD2}}{ for the case $\mathcal{T}=\mathcal{T}_c/2$, where $\mathcal{T}_c$ refers to the temperature where the perturbative instability sets in (which is smaller than the transition temperature in the first order case).}
\end{itemize}

\newpage

\end{document}